\documentclass{aa}

\input psfig.tex

\usepackage{graphicx}
\usepackage{natbib}
\usepackage{array}
\bibpunct{(}{)}{;}{a}{}{,}

\usepackage{rotating}       % for sideways tables/figures

\input psfig.tex
\usepackage{latexsym}
\usepackage{natbib}
\usepackage{amssymb}
\usepackage{amsmath}

\usepackage{graphicx}
\usepackage{graphics}
\usepackage{fancyhdr}
\usepackage{morefloats}

\begin{document}

\title{Formation and evolution of early-type galaxies.\\
II. Models with quasi-cosmological initial conditions }
\subtitle{}
     \author{  Emiliano Merlin \inst{1} \&
           Cesare Chiosi \inst{1}
            }
     \offprints{E. Merlin }
     \institute{$^1 $ Department of Astronomy, University of Padova,
                Vicolo dell'Osservatorio 2, 35122 Padova, Italy  \\
                \email{merlin@pd.astro.it; chiosi@pd.astro.it }
      }
     \date{Received: November 7, 2005; Accepted: April 12, 2006}

%=================== BEGIN ABSTRACT ===================%
\abstract {In this study, with the aid of \textit{N-Body}
simulations based on quasi-cosmological initial conditions, we have
followed the formation and evolution of two early-type galaxies of
different total mass, from their separation from global expansion of
the Universe to their collapse to virialized structures, the
formation of stars and subsequent nearly passive evolution. Using
the \textsc{PD-TSPH} N-Body code, we developed the simulations of
two model galaxies in quasi-cosmological context. The cosmological
background we have considered is the Standard CDM. The models are
made of Dark and Baryonic Matter in standard cosmological
proportions (9:1), and have significantly different initial total
mass, i.e. $1.62\times 10^{12} M_\odot$ (A) and $0.03\times 10^{12}
M_\odot$ (B). Particular care has been paid to the star formation
process, heating and cooling of gas, and chemical enrichment. Star
formation is completed within the first 3 Gyr in Model A, whereas it
lasts longer up to about 4 Gyr in Model B. The models are followed
for a long period of time, i.e. 13 Gyr (Model A) and 5 Gyr (Model
B): in any case, well beyond the stages of active star formation.
The structural properties of the present-day models are in good
agreement with current observations. The chemical properties, mean
metallicity and metallicity gradients also agree with available
observational data. Finally, conspicuous galactic winds are found to
occur. The models conform to the so-called \textit{revised
monolithic scheme}, because mergers of substructures have occurred
very early in the galaxy life. Our results agree with those obtained
in other similar recent studies, thus strengthening the idea that
the revised monolithic scheme is the right trail to follow in the
forest of galaxy formation and evolution. }

\titlerunning{Formation and evolution of early-type galaxies}
\authorrunning{E. Merlin \& C. Chiosi }
\maketitle

%=================== BEGIN DOCUMENT ===================%

\section{Introduction}\label{intro}

One of the major challenges in modern astrophysics is to understand
the origin and evolution of galaxies, the bright elliptical ones in
particular. In a Universe dominated by Cold Dark Matter (CDM) and,
as recent observations seem to suggest, by some kind of Dark Energy
in the form of a non-zero Cosmological Constant $\Lambda$, and
containing a suitable mix of baryons and photons, cosmic structures
are formed by the gravitational collapse of Dark Matter and are
organized in a hierarchy of complexes (halos) inside which baryons
dissipate their energy and collapse to form luminous systems. In
this context, modelling the formation of early-type galaxies with
simulations taking into account the dynamics of Dark Matter halos
and gas, radiative cooling of baryons, star formation, and gas loss
by galactic winds can be reduced to following schemes
\citep{Peebles02,Schade99}.

(1) The \textit{monolithic} scenario of galaxy formation supposes
and predicts that all early-type galaxies form at high redshift by
rapid collapse and undergo a single prominent star formation episode
\citep[e.g.][]{Eggen62,Larson75,Arimoto87,Bressan94} ever since
followed by quiescence. In favour of this scheme are the
observational data that convincingly hint for old and homogeneous
stellar populations \citep[see][for a recent review of the
subject]{Chiosi00}. It is worth mentioning, however, that
\citet{Kaufmann93} and \citet{Barger99} argue for the presence of
recent evolution in the stellar populations of elliptical galaxies.

(2) The \textit{hierarchical} scenario suggests instead that massive
galaxies are the end product of subsequent mergers of smaller
subunits, over time scales almost equal to the Hubble time in the
particular cosmological model used; as the \textit{look back time}
increases, the density in comoving space of bright (massive)
elliptical galaxies should decrease by a factor 2 to 3
\citep[e.g.][]{White78,Kaufmann93}. In favour of this view are some
observational evidences that the merger rate likely increases with
$\sim (1+z)^3$ \citep{Patton97} together with some hint for a
color-structure relationship for E \& S0 galaxies: the color becomes
bluer at increasing complexity of a galaxy structure. This could
indicate some star formation associated to the merger event.
Finally, there are the many successful  numerical simulations of
galaxy encounters, mergers and interactions
\citep[e.g.][]{Barnes96}. Nevertheless, contrary to the expectation
from this model, the number density of elliptical does not seem to
decrease with the redshift, at least up to  $z \simeq 1$
\citep{Im96}.

(3) There is a third scheme named \textit{dry merger view}, in which
bright ellipticals form by encounters of quiescent, no star forming
galaxies. This view is advocated by \citet{Bell04} who finds that
the \textit{B}-band luminosity density of the red peak in the colour
distribution of galaxies shows mild evolution starting from $z
\simeq 1$. As old stellar populations would fade by a factor 2 or 3
in this time interval, and the red colour of the peak tells us that
new stars are not being formed in old galaxies,  he argues that this
mild evolution hints for a growth in the stellar mass of the red
sequence, either coming from the blue-peak galaxies in which star
formation is truncated by some physical process, or by ``dry
mergers'' of smaller red, gas-poor galaxies.

(4) Finally, a fourth hybrid scenario named \textit{revised
monolithic} has been proposed by \citet{Schade99}, who suggests that
a great deal of the stars in massive galaxies are formed very
early-on at high $z$ and the remaining few ones at lower $z$. The
revised monolithic ought to be preferred to the classical
monolithic, as some evidences of star formation at $0.2 \leq z \leq
2$  can be inferred from the presence of the emission line of [OII],
and also as the number frequency of early-type galaxies up to $z
\simeq 1$ seems to be nearly constant.

However, a sharp distinction among the various scenarios could not
exist in reality. As pointed out by \citet{Longhetti00} early-type
galaxies in isolation and in interaction (such as pair and shell
galaxies) share the same distribution in  diagnostic planes such as
$H{\beta}$ vs [MgFe], the classical tool to infer age differences.
This means that secondary episodes of star formation may not only
occur in merging (dynamically interacting) galaxies, but also in the
isolated ones because of internal processes,  thus supporting the
Schade \textit{revised monolithic} scheme. Captures (mergers) of
small satellites by a galaxy born in isolation according to the
monolithic scheme are reasonably possible (e.g. the Milky Way which
is currently capturing the Sagittarius dwarf galaxy). In any case,
as thoroughly discussed by \citet{Chiosi00},
\citet{ChiosiCarraro02}, \citet{Tantalo04a} and \citet{Tantalo04b},
the age and intensity of the last episode of star formation
(measured by the fractionary mass engaged in newly formed stars)
cannot exceed some stringent limits, a few per cent and about one
third of the Hubble time even in a massive elliptical, otherwise the
typical broad band colours cannot be matched; they would be too blue
compared to observations.

In addition to all this, elliptical galaxies are known to obey to a
colour-magnitude relation \citep{Bower92}: the colours become redder
at increasing luminosity (and hence mass) of galaxies. The relation
is tight for cluster galaxies and more disperse for field galaxies.
The colour of a galaxy simply mirrors the colours of the constituent
stellar populations. It can be shown that the colour-magnitude
relation stems from  a mass-metallicity sequence and not an age
sequence, see for instance \citet{Kodama97} and
\citet{ChiosiCarraro02} and references therein.

Long ago, \citet{Larson75} advocated the supernova driven galactic
wind model to explain the colour-magnitude relation. He argued that
massive galaxies retain gas and form stars for longer periods of
time, and become richer in metals than the low-mass ones. In any
case the star forming period even in the most massive galaxies
seemed to be confined at very early epochs (say within the first Gyr
to be generous). This kind of model has been extensively used to
describe and predict the chemo-spectro-photometric properties of
elliptical galaxies
\citep[e.g.][]{Bressan94,Bressan96,Tantalo96,Tantalo98,Chiosi98}.
The short duration of the dominant star forming period has also been
suggested by \citet{Bower92} on the basis of  very simple, empirical
arguments: the vast majority of stars are formed at $z > 2$ and
subsequent activity if any should have occurred at $z \geq 0.5$.

Over the years, the situations became more intrigued because (i) the
\citet{Larson75} model got in contrast with the observational
constraint that massive elliptical have on the average [$\alpha$/Fe]
ratios higher (super-solar) than the low mass ones (solar or below
solar), thus implying the opposite trend for the duration of the
star forming period as amply discussed by \citet{Chiosi98},
\citet{Chiosi00} and \citet{ChiosiCarraro02}; (ii) of the results
from absorption line indices, which seem to indicate that large age
difference could exist. According to \citet{Bressan96},
\citet{Tantalo98}, \citet{Tantalo02}, \citet{Tantalo04a} and
\citet{Tantalo04b}, the following picture emerges: all elliptical
galaxies have been formed far back in the past, but have undergone
different histories of star formation. Massive ellipticals had a
single burst-like episode of stellar activity ever since  followed
by passive evolution, whereas the low mass ones stretched their star
formation history over long periods of time in a series of bursts.
This view has been supported by the N-Body simulations of
\citet{ChiosiCarraro02} who find that the duration and mode of star
formation is driven by the total mass (Dark plus Baryonic Matter)
and/or the initial density of the galaxy. In addition to this the
\citet{ChiosiCarraro02} models while predicting copious galactic
winds much improved upon the simple-minded scheme by
\citet{Larson75} and ruled out the point of inconsistency with the
abundance ratios. We will came back to these issue later on. It is
also supported by the most recent Spitzer-data (and SIRTF) in the
far infrared, which have revealed the existence of very massive
galaxies already in place at redshift $z \sim 6$; in addition to
data HST observations have also brought into evidence galaxies in
place at $z \sim 7-8$ \citep{Bouwens04}.

All this can be understood within a model ruled by \textit{violent
relaxation} taking place in the remote past thanks to which clusters
of just born stars can merge together into even denser clusters
\citep[e.g.][]{Macchetto96}.  In this context it is also worth
mentioning that according to \citet{Bundy05} (i) the mass of star
forming galaxies increases with redshift (\textit{"downsizing"})
thus strongly weakening the hierarchical scenario as the main
mechanism for forming  giant ellipticals; and (ii) even if "dry
mergers" clearly occur \citep{vanDokkum05}, they cannot play an
important role in the assembling history of massive quiescent
galaxies as the expected number density of massive galaxies at low
redshift does not agree with observational data. Furthermore  the
mass-downsizing of star forming galaxies seems to be almost
independent of the environmental density, whereas one would expect a
strong density dependence in the dry merger rate. Therefore, either
classical or revised monolithic scenarios seem to be the right frame
for the galaxy formation mechanism \citep{Peebles02,Schade99}.

Basing on the pioneer studies by \citet{Katz91} and
\citet{KatzGunn91},  and later by many authors among whom we recall
\citet{Kawata99,Kawata01b,Kawata01a,Kawata03} and
\citet{Kobayashi05}, we present  numerical simulations carried out
with the \textsc{PD-TSPH} code \citep{Carraro98} in order to
investigate the formation mechanism on galactic scales. The scenario
we have in mind and we intend to prove with the present simulations
is the following one. At a certain (high) redshift a massive
perturbation made of Dark Matter and baryons detaches from the
Hubble flow and collapses on its own. It becomes a massive
proto-galaxy, rich of substructures which moving inside the common
gravitational potential well merge, form stars and eventually give
rise to a single entity (the galaxy). The process is complete at
high redshift (say before $z\simeq 2$) and from now on the galaxy
evolves in "isolation"; subsequent captures of small satellites are
possible without significantly altering the overall structure and
evolution. In the case of a low-mass perturbation the stellar
activity is prolonged over long time scales. Mergers between two
galaxies may occur but they are a sort of rare, spectacular event
and not the basic mechanism for assembling massive ellipticals.

The plan of the paper is as follows.  Section \ref{ini_cond}
presents the procedures we have used to derive   the initial
conditions  for our models; the procedure we have adopted is
outlined and compared with other techniques available in literature.
Section \ref{basic_phys} describes some assumptions we have made for
some fundamental physical processes that are included in the
numerical simulations, namely cooling, star formation and initial
mass functions, heating and chemical enrichment, and shortly
outlines the numerical implementation of these processes in the
N-Body PD-TSPH code developed by the Padova group over the years,
together with a brief description of the probabilistic view of star
formation, heating, cooling and chemical enrichment proposed by
\citet{Lia02}. Section \ref{two_models} presents the key parameters
of the models we plan to calculate, and in particular it deals with
the initial conditions derived from the method described in Section
\ref{ini_cond}. Section \ref{results} presents the results: the
spatial evolution, the star formation history, the energy balance,
the chemical enrichment and other details. Finally, Section
\ref{conclusions} contains  some concluding remarks.

\section{Search of the initial conditions } \label{ini_cond}

The choice of the initial conditions is perhaps the fundamental step
on which much of the future evolution of structures (galaxies) will
depend. As noted by \citet{Power03}, despite the popularity of
cosmological N-body simulations in the last years, there is very
little detail in literature about setting up initial conditions.

The scheme we have adopted can be defined as
\textit{``quasi-cosmological''}.  It starts from realistic
simulations of the cosmological evolution of the Universe that are
suitably adapted to the numerical simulations of galaxy formation
and evolution.

\subsection{Quasi-cosmological initial conditions}  \label{semicosmo}

Our initial conditions stem from the studies of \citet{Katz91},
\citet{KatzGunn91}  and \citet{Kawata99}. We adopt the Standard Cold
Dark Matter Universe  with $\Omega(z)=\Omega_0$, the matter density
at the present time. Therefore, the basic parameters of the model
are $H_0$, $\Omega_0$, and $\sigma_8$ (the rms mass fluctuations of
the present-day Universe that normalizes the density fluctuations
and fix the initial redshift of the simulations via the amplitude of
the density fluctuations)\footnote{All equations used in this
section are from \citet{Pad93} unless otherwise specified.}. These
models do not consider the presence of the Cosmological Constant
$\Lambda$ as indicated by the recent W-MAP data \citep{Spergel03}.

\subsubsection{Preparing the quasi-cosmological models}\label{semicosmo_mod}

To derive the initial conditions (positions and velocity field) of
Dark and Baryonic Matter particles of the pro-galaxy, we start from
simulating an initial large grid representing the Universe using the
public software \textsc{Grafic2} by \citet{Bertschinger01}. The
initial Universe grid is a very large cube with $\sim 40$ comoving
Mpc per side. In the grid we select a peak of over-density which is
assumed as the centre (or very close to it) of a new cubic sub-grid
of smaller size. In practice, we fix the size of the sub-grid
looking for the distance from the peak at which the over-density
goes to zero. This distance is  taken  as half the size of the cube
dimension $l$. In this sub-grid both Dark and Baryonic Matter are
present. Their respective barycentres  do not necessarily coincide
each other nor are both coincident with the geometrical centre of
the cube. Then we isolate this cubic volume, and using
\textsc{Grafic2} multi-level initial conditions generator we improve
upon its resolution, thus describing small-scale perturbations in
great detail, superposed to the large scale perturbations described
in the first-level grid. Subsequently, using the radial distance of
each particle (Dark Matter and Baryons) from the centre of the cube,
we select the sample of particles contained in a sphere with radius
equal to $l/2$. The volume of the sphere is about half of the volume
of the cube and accordingly the number of particles in the sphere is
about half of the number of particles in the cube. It is worth
recalling that, due to the cosmological perturbations, as  the
baryons in the cube could have a slight offset with respect to Dark
Matter, the same could happen with the spheres of baryons and Dark
 matter. Therefore, the numbers of
particles of the two matter species can  be slightly different, and
 not be exactly concentric. This however is less of  a problem
 because first it is physically understandable (perturbations of the
density field and  infall of baryons in the Dark Matter potential
wells), and second the difference in the numbers of particles is
always small. By replacing the cube with a sphere  we avoid spurious
tidal effects on the particles.  The sphere represents our
proto-galaxy in very early stages. It is worth noticing that
centering the new coordinate system at the geometrical centre of the
cube rather than at the mass-weighed centre leads to the possibility
that the peak of the density perturbation does not coincide with the
centre of the sphere. The particle positions are then referred to
the new system: the distances provided by \textsc{Grafic2} in
\textit{comoving} Mpc are translated to proper distances in Mpc
according to

\begin{equation}
d_{pr} =a_{start} \times d_{comov}
\end{equation}

\noindent  where $a_{start}=(1+z_{start})^{-1}$ is the value of the
expansion factor at the initial  redshift, $d_{comov}$ is the
distance in the comoving frame of reference and $d_{pr}$ is the
proper distance. The velocities are then referred to the centre of
the sphere, dropping the motion of the sphere as a whole and leaving
to each particle only the peculiar motion. To each particle we add
the velocity of the Hubble flow, which is simply given by

\begin{equation}
v_{flow}=H(z_{start}) \times d_{pr}
\end{equation}

\noindent (this velocity is radially oriented). Note that in order
to calculate $H(z)$ one has to assume a cosmological model. In the
Standard CDM cosmology

\begin{equation}
H(z)=H_0 \sqrt{\Omega_m (1+z)^3 }.
\end{equation}

\noindent

Finally we add a certain amount of rigid rotation. The associated
spin parameter $\lambda$ is

\begin{equation}
\lambda = \frac{J |E|^{1/2}}{G M^{5/2}}
\end{equation}

\noindent where $J$ is the angular momentum, $E$ is the  initial
binding energy,  and $M$ the total mass of the system. Typical
values of $\lambda$ range from 0.02 and 0.08 \citep{White84}, which
corresponds to angular velocities of the order of fractions of a
complete rotation over time-scales as long as about ten
\textit{free-fall} time-scales \citep{Carraro98}. As the
\textit{free-fall} time-scale depends on the density according to

\begin{equation}
t_{ff} = \frac{1}{\sqrt{4 \pi G \rho}}
\end{equation}

\noindent it strongly depends also on the initial redshift. The
background  density for matter scales indeed as $(1+z)^3$.
Introducing the rigid rotation somehow simulates and takes into
account the effects of fluctuations with wavelength longer than the
typical size of the proto-galaxy \citep[e.g.][]{Kawata99};
neglecting rotation would imply that the matter surrounding the
proto-galaxy has no effect on its evolution.

\subsubsection{Densities and masses: some cautionary remarks}\label{semicosmo_mass_den}

The theory of spherical collapse \citep[cfr.][]{Pad93} would allow
us to estimate the approximate redshift at which turn-around and
collapse of an homogeneous and expanding sphere should occur under
the action of self-gravitation. The relations of interest here are

\begin{eqnarray}
1+z_{turn} = 0.57(1+z_i) \delta_i  \nonumber \\
1+z_{coll}\,\, = 0.36(1+z_i) \delta_i
\end{eqnarray}

\noindent where $\delta_i$ is the mean over-density of the sphere
(proto-galaxy) at the initial redshift. But these values are mere
approximations of reality.

First of all, the classical theory of spherical collapse stands on
\textit{top-hat} perturbations, in which the particle velocity field
is determined only by the Hubble flow. The presence of peculiar
velocities and density perturbations strongly alters the predictions
of this simple theory.  If the initial conditions are such that
there is a single, strong peak in the density field, a more reliable
model could be the \textit{secondary infall model}
\citep[e.g.][]{Miller86}. Models for the spherical collapse taking
the initial kinetic energy of peculiar motions into account do
actually exist, see e.g. \citet{Pad93}. As these models go  beyond
our aims, they will not be consider here. Second, cutting the sphere
out of the cosmological Universe-grid and isolating it in a void
space implies that the mean density of the  Universe coincides with
the mean density of the sphere.  Therefore, assigning a certain
over-density to the sphere will force it to follow an evolutionary
path typical of a Universe with $\Omega>$1. This means that it will
collapse on a faster time-scale. In cosmological simulations, the
problem is nowadays  solved using \textit{periodic boundary
conditions}. To cope with this point of uncertainty in our models,
we have limited ourselves to consider only
 low  initial over-densities ($\leq 0.1$).

Furthermore, even if an univocal relationship should exist between
redshift and proto-galaxy over-density, in reality fluctuations
around the mean value of  $\delta$ are possible, the fluctuations
being larger at decreasing mass of the proto-galaxy.

Finally, the spherical symmetry and sharp confinement of matter in
the sphere we have adopted to avoid complications due to possible
irregular shapes of the proto-galaxy, imply that particles in the
outskirts of the galaxy will meet sudden vanishing of density and
velocity at the surface of the sphere.

Given all these uncertainties, it is clear that the results of the
simulations will be only loosely related to  theoretical predictions
such as expected collapse redshifts.

Considering the total mass as a free parameter would allow to
investigate a wider range of possibilities, but such a choice would
somehow weaken the above quasi-cosmological scheme and the peculiar
velocities (which depend on the overdensity) should be properly
adjusted. Therefore we  prefer to assign the sphere the mass given
by the "correct" value of the mean overdensity output by
\textsc{Grafic2}.

\subsection{Comparison with other initial conditions }\label{comp_ini_cond}

It is worth commenting here the main difference between our initial
conditions and those adopted by \citet{Carraro98} and
\citet{ChiosiCarraro02}. They start with a spherical model
containing  Dark and Baryonic Matter with masses in the standard
proportions  9:1. They set the initial density profile for the Dark
Matter component to be

\begin{equation}
\rho(r) = \rho_c \frac{r_c}{r} \label{navar}
\end{equation}

\noindent where $r_c$ and $ \rho_c$ are radius and density of a
small central  sphere. It resembles the well-known \citet{NFW96} law
(see below) in the central regions. This indeed, independently of
cosmological models, initial spectrum of perturbations and total
mass, this profile is $\propto r^{-1} $ in the central regions and
$\propto r^{-3} $ in the external ones. The spatial positions of
Dark Matter particles are obtained by MonteCarlo deviations from the
distribution law. The initial velocities of Dark Matter particles
are obtained from the velocity dispersion $\sigma (r)$  assuming
equi-partition among the three components, i.e.

\begin{equation}
v(r) = \frac{1}{3} \sigma (r) = \sqrt{\rho_c r_c G r
\ln(\frac{R_T}{r})}
\end{equation}

\noindent which is obtained by inserting eqn. (\ref{navar}) in

\begin{equation}
\rho(r) \sigma(r)^2 = \int_r^{R_T} \frac{G M(r')}{r'^2} \rho(r') dr'
\end{equation}

\noindent \citep{Binney87}.

The gas particles (baryons) are then homogeneously and randomly
distributed in the spherical halo of cold Dark Matter with null
velocity field to simulate the \textit{infall} of primordial gas
into the potential well of Dark Matter \citep{White78}.

In practice, this is equivalent to start with a halo of Dark Matter
already detached from the Hubble flow which begins to collapse
carrying along the baryonic matter. Even if all this sounds
physically reasonable, these initial conditions already contain the
solution of the problem: a self-gravitating, collapsing halo of Dark
Matter whose density profile already  resembles the final structure
one is looking at. Moreover, in the CDM cosmology  catching up of
baryons by the potential wells of Dark Matter happens  much earlier
than the time at which a proto-galactic halo detaches from the
Hubble flow.

The initial radius is set to be equal to the \textit{virial radius}
$R_{200}$ of the proto-galaxies, which is defined as the radius of a
sphere with mean density equal to 200 times the mean background
density of the Universe, $\rho_u$, which in the case of a Standard
CDM model ($\Omega_0$ = 1) is given by

\begin{equation}
\rho_u(z)=\frac{3H_0^2}{8\pi G}(1+z)^3
\end{equation}

\noindent with obvious meaning of the symbols, since the minimum
density  required for collapse is about $200 \rho_u(z)$
\citep{NFW96,Pad93}. Finally, star formation is supposed to occur
only after the virialization  of the CDM halo.

To summarize, the present initial conditions differ from those used
in past similar studies carried out by the Padua group with the same
software \citep[e.g.][]{ChiosiCarraro02} at least in the three
aspects:

\begin{itemize}
\item The initial velocities taking into account the peculiar motions
and the Hubble flow provide a more realistic description of the
complexity of the galaxy formation process as compared to the
"static" model.

\item The variation of the perturbation amplitude and spectrum
allows us to better explore their
effects on the resulting galaxy models and properties.

\item The peculiar velocities are set to their initial value in
a self-consistent way securing the correct growth of the
perturbations on all scales of interest \citep[e.g.][]{Katz91}.
\end{itemize}

Modern numerical cosmological simulations include more sophisticated
methods, such as evolution in comoving space and
\textit{resimulation} at different degrees of accuracy and
resolution \citep[see, e.g.][]{Kawata03}. The present scheme has to
be intended as the initial step towards future improvements.
Nevertheless we are confident that the present method, despite its
limitations, is reasonably suited to study  the formation and the
morphological, dynamical and chemical evolution of early-type
galaxies as isolated objects.

\section{Basic physics in galaxy models}\label{basic_phys}

 In this section we shortly summarize the basic physical processes that are
thought to lead  galactic formation and evolution, together with
some details on their implementation in the fully lagrangian
\textsc{PD-TSPH}. More details on the code can be found in
\citet{Carraro98}, \citet{Buonomo00}, \citet{Lia02} and
\citet{ChiosiCarraro02}.

Key physical ingredients ingredients of the whole problem  are the
gravitational force that acts on all kinds of particle and is
implemented by the classical \textit{Tree-code} of
\citet{BarnesHut86} with quadrupole expansion, and the hydrodynamic
processes, which instead act only on the energy and motions of the
gas particles together with star formation, chemical evolution, gas
cooling by radiative processes, and finally gas heating by energy
injection from supernova explosions, stellar wind, UV cosmic
radiation and others.

The system of gas particles can be viewed as a collisional fluid governed
by the classical equations, which in lagrangian notation are

\begin{center}
\begin{eqnarray}
\frac{d \rho}{dt} &=& - \rho \nabla \overrightarrow{v}        \label{cont} \\
\frac{dv}{dt}     &=& -\frac{1}{\rho} \nabla P - \nabla \Phi  \label{eul} \\
\frac{du}{dt}     &=& - \frac{P}{\rho} \nabla u + S           \label{ene} \\
\nabla ^2 \Phi    &=& 4 \pi G \rho \label{pois}               \label{poisson}
\end{eqnarray}
\end{center}

\noindent i.e. the \textit{continuity} or mass conservation
equation, the \textit{Euler} or momentum conservation equation, the
\textit{energy} conservation equation, and the \textit{Poisson}
equation linking the gravitational potential to the medium density,
respectively. The companion equation of state is that of a perfect
gas $P = \frac{R \rho T}{\mu}$.

Gas-dynamics is described with the \textit{SPH} formalism by
\citet{Monaghan85} and \citet{HernquistKatz89}, as implemented by
\citet{Carraro98}, \citet{Buonomo00}, \citet{Lia02} and
\citet{ChiosiCarraro02}; in particular, a probabilistic approach to
non-adiabatic processes such as star formation and gas restitution
is used (see below). The term $S$ in eqn.(\ref{ene}) is the energy
source (otherwise known as \textit{source function}) from all
\textit{heating} and \textit{cooling} processes. It is written as

\begin{equation}
S = \frac{\Lambda - \Gamma }{\rho}
\end{equation}

\noindent where  $\Lambda$ and $\Gamma$ are the cooling and heating
rates,  respectively (see \ref{cool} and \ref{heat} for details).

To derive the mean value of any physical quantity $f(r)$ over a
certain interval we adopt the so-called \textit{``gather/scatter''}
functional  relation suggested by \citet{HernquistKatz89}:

\begin{equation}
<f(r)>= \int f(r') \frac{1}{2} \{W(r-r',h(r'))+W(r-r',h(r))\} dr',
\end{equation}

\noindent
 where $h$ is the \textit{smoothing length}, which gives
the extension of the volumes over which a suitable space average is
made.  For the \textit{kernel} W(r) we adopt the form proposed by
\citet{Monaghan85}

\begin{equation}
W(r,h)=\frac{1}{\pi h^3} \left\{
\begin{array}{ll}
1-\frac{3}{2} (\frac{r}{h})^2 + \frac{3}{4} (\frac{r}{h})^3 \,\, \mbox{ if }
0 \leq \frac{r}{h} \leq 1 \\
\frac{1}{4} (2-(\frac{r}{h}))^3 \,\, \mbox{ if } 1 \leq  \frac{r}{h} \leq 2 \\
0 \,\, \mbox{ if }  \frac{r}{h} \geq 2.
\end{array}
\right.
\end{equation}

As usual, in the Euler equation (\ref{eul}) an additional viscosity
term is introduced to describe more complicated processes than
simple shock waves. There are many possible formulations for this
additional term, among which we adopt the one  by \citet{Monaghan85}

\begin{equation}
\prod_{ij} = \frac{ -\alpha \mu_{ij} \overline{c_{ij}} +  \beta
\mu_{ij}^2}{\overline{\rho_{ij}}}\overline{f_{ij}}
\end{equation}

\noindent
where

\begin{equation}
\mu_{ij} = \left\{
\begin{array}{ll}
\frac{v_{ij} r_{ij}}{h_{ij} (r_{ij}^2 / h_{ij}^2 + \eta^2)}
\,\,\,\mbox{ if } v_{ij} r_{ij} < 0 \\
0 \,\,\,\mbox{ if } v_{ij} r_{ij} \geq 0
\end{array}
\right.
\end{equation}

\noindent
 with the following meaning of the symbols:  $v_{ij} = v_j
- v_i, \overline{c}_{ij} = (c_j + c_i)/2$ is the mean sound
velocity; $h_{ij} = (h_j + h_i)/2, \overline{\rho}_{ij} = (\rho_j +
\rho_i)/2 $,  $\alpha, \beta$ and $\eta$ are constant coefficients
with typical values   $ \sim 1, \sim 2, and \sim 0.1 $, respectively
(the latter coefficient is inserted only to prevent numerical
divergences); and $\overline{f_{ij}}$ represents a shear correcting
term to the artificial viscosity \citep[cfr.][]{Steinmetz96}.

The  code calculates the  \textit{time-step} of each individual gas
particle by means of the \textit{Courant condition},

\begin{equation}
\Delta t_{C,i} = \frac{C h_i}{[h_i |\nabla v_i| + c_i + 1.2 ( \alpha
c_i + \beta_max_j | \mu_{ij}|)]}
\end{equation}

\noindent where C is a constant nearly equal to $\sim 0.3$. For all
kinds of particle (both collisional and collisionless)  in presence
of gravity a more stringent condition is required
\citep{KatzWeinberg96}

\begin{equation}
\Delta t_{G,i} = \eta \times {\rm min} \left[ \frac{\eta
\epsilon}{|v|}, \, \left(\frac{\epsilon}{|a|}\right)^{1/2}\right]
\end{equation}

\noindent where  $\epsilon$ is the \textit{softening}  parameter and
$\eta$ is constant with value  $\sim 0.5$. Finally, the
\textit{time-step} for the i-th particle is the least of the two:
$\Delta t_i = {\rm min} ( \Delta t_{C,i}, \Delta t_{G,i})$; and the
time  step for the whole system is taken to be the smallest of all
particles time steps.

Furthermore, the \textit{smoothing length} is varied with time and
space  according to the prescription by  \citet{Benz90}

\begin{equation}
\frac{dh}{dt} = -\frac{1}{3} h \nabla v.
\end{equation}

\noindent  The softening parameters we have adopted for the
different components of our simulations are given in Table
\ref{tabdyn}.

\subsection{Cooling} \label{cool}

Gas \textit{cooling} by many radiative processes perhaps plays the
dominant  role in the collapse of baryonic matter in galaxies and
star formation, since it is the comparison between the cooling and
the dynamical time scales, respectively given by

\begin{displaymath}
t_{cool} = \frac{E}{\dot{E}} \simeq \frac{3 \rho k T}{2 \mu  \Lambda
(T)} \quad {\rm and} \quad t_{dyn} \simeq \frac{\pi}{2}
\sqrt{\frac{2GM}{R^3}}
\end{displaymath}

\noindent (where $\rho$ is the mean gas density and $\Lambda (T)$ is
the  cooling rate of the gas at the temperature T) that decides
whether the gas will collapse nearly on the \textit{free-fall} time
(if $t_ {cool} < t_{dyn}$) or just contract through subsequent
equilibrium conditions (if $t_ {cool} > t_{dyn}$).

Many cooling processes are known. In general, the \textit{radiative
cooling}  depends on density, chemical composition and temperature
(from this latter very strongly). For $ T > 10^4 K$ the most
efficient process is \textit{bremsstrahlung emission} from the
ionised plasma. In the range $100 \leq T \leq 10^4 K$ collision
between H$_2$ molecule and/or H atoms dominate via rotational and
vibrational energy decay. Finally, for $ T < 100 K$, the dominant
contribution comes from CO molecules. See \citet{Chiosi98} for more
details on the subject.

Another important cooling mechanism is the \textit{inverse Compton
effect}, which is particularly important at high redshifts. The
\textit{cooling rate} of a gas with electron number density $n_e$
and temperature $T$, in a field of radiation with density $\rho_R$
and temperature $T_R$, is

\begin{equation}
\Lambda_{comp}= \frac{4 \sigma_T n_e \rho_R (T - T_R)}{m_e}.
\end{equation}

\noindent
Estimating the \textit{cooling time} for matter immersed in the CMB we obtain

\begin{equation}
t_{comp}={ [3 m_p m_e (1+z)^{-4}] \over  (8 \mu \sigma_T \Omega_R \rho_{crit} ) }
\end{equation}

\noindent which is nearly equal to the dynamical time for collapse
red-shifts greater than 7, independently of the galaxy mass. As our
simulations begin at high redshift, the \textit{inverse Compton
effect} has been taken into account. In this paper we have adopted
the cooling rates given by \citet[][and references
therein]{Chiosi98} who have amalgamated results from different
sources and to whom the reader should refer for all details. The
same cooling rate have been adopted by \citet{Carraro98},
\citet{Buonomo00}, \citet{Lia02}, and \citet{ChiosiCarraro02} in
their studies of galaxy formation and evolution.

\subsection{Star formation rate and initial mass function } \label{sfr}

Star formation is among the most complicate and poorly known
physical aspects of astrophysics despite its relevance in many
issues. From simple minded arguments, there are at least three
prerequisites for a gas (molecular cloud) to be eligible to star
formation: the gas has to be in convergent motion, i.e. the velocity
divergence must be negative; the gas must be gravitationally
unstable, i.e. it must satisfy the Jeans condition $ \tau_{sound}
\geq \tau_{ff} $ (where $\tau_{sound}$ is the local sound velocity);
the gas must be cooling, i.e. it has to verify the relation $
\tau_{cool} \ll \tau_{ff} $ (see Section \ref{cool}). Then, the rate
of star formation (SFR) is customarily expressed by the
\citet{Schmidt59} law

\begin{equation}
\frac{d\rho_*}{dt_g} = -\frac{d\rho_g}{dt_g} = c^*\frac{\rho_g}{t_{g}} \label{schmidt}
\end{equation}

\noindent where $c^*$ is the so-called \textit{dimensionless
efficiency of star formation}, and $t_{g}$ is a characteristic time
scale. Normally, SPH codes treat star formation simply implementing
the Schmidt law in the computational language and transforming part
of gaseous particles which satisfy the three conditions above in
new, collisionless particles of different mass (``stars''), thus
giving rise to a huge increase in the total number of particles. In
order to avoid this creation of new particles with different mass as
a result of star formation, \citet{Lia02} proposed a new
interpretation of the Schmidt law: the star formation rate given by
eqn. (\ref{schmidt}) was interpreted as the probability that at each
time step a gas particle is instantaneously and fully turned into a
star particle (thus losing its collisional properties). They showed
that, averaged over a large fraction of the Hubble time, this
probabilistic view of the star formation rate converges to the
results of more conservative methods in which the masses of the
newly formed stellar particles strictly follow the dictates of the
chosen IMF law. The models we are going to present are based on this
view of the star forming process.

Furthermore,  even if \citet{Buonomo00} argued that the condition on
the velocity divergence of gas particles  (negative) is not strictly
required, the same arguments do not apply to our case because of the
initial expansion phase of our proto-galaxies. Neglecting this,
conditions, star formation could occur too early on.

Following \citet{Buonomo00}, we adopt  $c^*$=1.0 for all the models.
The characteristic time scale  is chosen to be the maximum between
$\tau_{cool}$ and $\tau_{ff}$ time-scales. Numerical experiments
show  that in most situations $t_g=\tau_{ff}$ is a good choice.

\citet{ChiosiCarraro02} have found that the star formation history
(rate vs time) depends on the depth of the gravitational potential
well of a galaxy. In the case of deep potentials (such as in massive
and/or dense galaxies), once star formation has started and energy
is injected to the gas by supernova explosions, stellar wind,
etc..., this is not enough to push the gas out of the potential
well. A sort of balance between cooling and heating is reached and
the gas consumption by star formation goes to completion. Star
formation cannot stop until the remaining gas is so little that any
further energy injection will eventually heat it up to such high
energies (temperatures) that the gravitational potential is
overwhelmed. No more gas is left over and star formation is
quenched. The star formation history resembles a strong unique burst
of activity, a sort of monolithic star forming event, taking place
over a certain amount of time, of the order of 1 to 2 Gyr. In
contrast, in a galaxy of low mass and/or density and hence shallow
gravitational potential, even a small star forming activity will
heat up the gas above the potential well. Some of it is soon lost in
galactic wind, the remaining one becomes so hot that it will take
long time to cool down and to form new stars. The cycle goes on many
times in a sort of repeated bursting mode of star formation taking
place during long periods of times if not for ever.

In each generation of stars, otherwise known as  \textit{Single
Stellar Population} (SSP), the stellar masses are known to obey a
distribution function, named \textit{Initial Mass Function} (IMF),
which is usually represented by a power law. The most popular of
these functions is the one by \citet{Salpeter55}. More recent ones
and currently used in galaxy evolution models are those by
\citet{Arimoto87} and \citet{Kroupa98}. The three IMFs are shortly
summarized here:

\begin{eqnarray}
\Phi_S(M) &=& C_S M^{-1.35} \quad {\rm Salpeter} \nonumber\\
\Phi_K(M)&=& \left\{
\begin{array}{ll}
C_{K1} M^{-0.5} \qquad \mbox{ if } M<0.5 \nonumber \\
C_{K} M^{-1.2}  \,\,\qquad \mbox{ if } 0.5<M<1   \quad {\rm Kroupa} \nonumber \\
C_{K} M^{-1.7}  \,\,\qquad \mbox{ if } M>1
\end{array}
\right. \\
\Phi_A(M) &=& C_A M^{-1.00}   \quad {\rm Arimoto \,\, \& \,\, Yoshii}
\end{eqnarray}

\noindent where $C_S $=0.1716,  $C_{K1}$=0.48, $C_K$=0.295  and
$C_A$=0.145. Needless to say, the IMF bears very much on many
aspects of galactic evolution. In our models we have adopted the
\citet{Kroupa98} IMF, originally designed to describe the situation
for the solar vicinity. We prefer not to use the \citet{Arimoto87}
IMF, even if it was claimed to be best suited to early-type
galaxies, because it would produce a too high metallicity (see
below).

%%%%%%%%%%%%%%%%%%%%%%%Table 1
\begin{table}
\caption{Initial parameters for Models A and B in the Standard CMD
cosmological background} \centering
\begin{tabular}{|l|l|l|}
\hline
Model                                   & A             & B       \\
\hline
Initial redshift                        & 50            & 53      \\
\hline
$\Omega_m$                              & 1             & 1       \\
\hline
$\Omega_\Lambda$                        & 0             & 0       \\
\hline
$H_0 = 50 \mbox{ } km Mpc^{-1} s^{-1} $ & 50            & 50      \\
\hline
$\sigma_8$                              & 0.5           & 0.5     \\
\hline
Spin parameter                          &  0.02         &   0.02  \\
\hline
Initial mean overdensity                & 0.12          &  0.12   \\
\hline
IMF                                     & Kroupa        & Kroupa  \\
\hline
\end{tabular}
\label{tabcosmo}
\end{table}

%%%%%%%%%%%%%%%%%%%%Table 2
\begin{table}
\centering \caption{\small{Initial dynamical and computational
parameters for Models A and B.}}
\begin{tabular*}{78mm}{|l|l|l|}
\hline
Model                                    & A     & B \\
\hline
Initial number of gas  particles         & 13719 & 13904 \\
\hline
Initial number of CDM particles          & 13685 & 13776 \\
\hline
Initial total mass ($10^{12}M_{\odot}$)  & 1.62  &  0.03 \\
\hline
Initial baryonic mass fraction           & 0.10  & 0.10 \\
\hline
Initial radius (kpc)                     & 33    & 9    \\
\hline
Softening parameter for gas (kpc)        & 1 & 0.5  \\
\hline
Softening parameter for  DM (kpc)        & 2 & 1 \\
\hline
\end{tabular*}
\label{tabdyn}
\end{table}

\subsection{Heating} \label{heat}

Gas heating is due to supernova explosions, stellar winds  and UV
radiation emitted by massive stars, cosmic ultraviolet light, and to
the cosmic background radiation. In addition to this we have several
mechanisms of mechanical and dynamical nature which are already
accounted for by the first term of the energy in eqn. (\ref{ene}).

The total heating rate by radiative processes is given by

\begin{equation}
\Gamma = \frac{E_{SNI} + E_{SNII} + E_w + E_{UV}}{\Delta t}
\end{equation}

\noindent where the terms refer to Type Ia and Type II supernovae,
stellar winds and ultraviolet flux, respectively. In our model we
consider only the supernova heating and neglect the other two. In
brief, (i) the UV flux from massive stars is absorbed by the
interstellar molecular gas and re-emitted in the far infrared and
thus is soon lost by radiation; (ii) the kinetic contribution by
stellar wind should be treated on much smaller scales than those
permitted by the resolution of these simulations. This is a point to
keep in mind because the energy budget from stellar winds from
massive stars may parallel that by supernova explosions.  Finally,
we do not include the heating by UV cosmic radiation. This may be a
point of weakness because this heating source could delay gas
cooling  and  star formation in turn, and even stop star formation
at later times by contrasting gas cooling. Although according to
\citet{Navarro97} it can be neglected at least during the first
stages of galaxy formation, careful investigation of this issue is
mandatory to properly understand the mechanisms of star formation.

Supernova explosions return nuclearly processed gas to the
interstellar medium, and heat it up. Type Ia supernovae from binary
systems and Type II supernovae from massive stars affect in a
different way the heating of the interstellar medium because of
their much different evolutionary time scales. The progenitors of
Type Ia are CO-white dwarf accreting mass from a companion, which in
turn can be either a red giant or another CO-white dwarf. When and
if the mass accreting white dwarf grows to the Chandrasekhar limit
1.4 $M_\odot$,  C-ignition can take place inducing thermal runaway
and consequent supernova explosion. The time scale for this to occur
ranges from a few hundreds Myr to several Gyr. In contrast, Type II
supernovae generated by core collapse in massive stars have time
scales of the order of a few Myr. Taking into account such different
behaviour is an important aspect of modelling galaxy evolution. We
assume that each supernova explosion liberates $k_{SN} \times
10^{51}$ erg of energy; anyway, only a small fraction of which - say
$10\%$ - can be thought as given back to the ISM \citep{Thornton98};
the parameter $k_{SN}$ fixes the amount of energy which is actually
made available to the gas and not dispersed by radiative processes.
This energy injection increases the internal energy of the gas
particles in the proximity of the explosion according to the SPH
formalism and some kinematical / dynamical effects ought to be
expected. However, as for stellar winds, it is difficult to
correctly take into account the effect of this on the kinematics of
the nearby gas particles, as the scales on which these effects
should be noticeable are too small if compared to the resolution of
our simulations. Therefore we simply give back the whole amount of
energy to the thermal budget of the nearby particles. This
approximation has little impact on the dynamical  evolution of the
whole system, due to the high cooling rates involved in the process
\citep[see e.g.][]{Carraro98,Kawata03}.

To describe the rates of supernova  explosions we adopt the popular
formulation by \citet{Greggio83}; see there and references therein
for details. To describe the energy feedback, we adopt once more the
probabilistic description of \citet{Lia02}, to which we remand for a
detailed description.

%%%%%%%%%%%%%%%%%%%%%%%%FIG 1
\begin{figure}
\centering
\includegraphics[width=7cm,height=7cm]{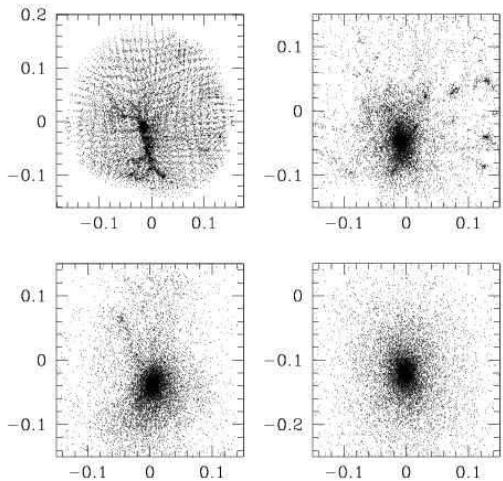}
\caption{Positions on the $xy$ plane of the Dark Matter particles
for Model A (proper Mpc) at different times; from left to right and
top to bottom, $z= 7.6, 3.2, 1.6, 0$.} \label{pos1}
\end{figure}

%%%%%%%%%%%%%%%%%%%%%%%%FIG 2

\begin{figure}
\centering
\includegraphics[width=7cm,height=7cm]{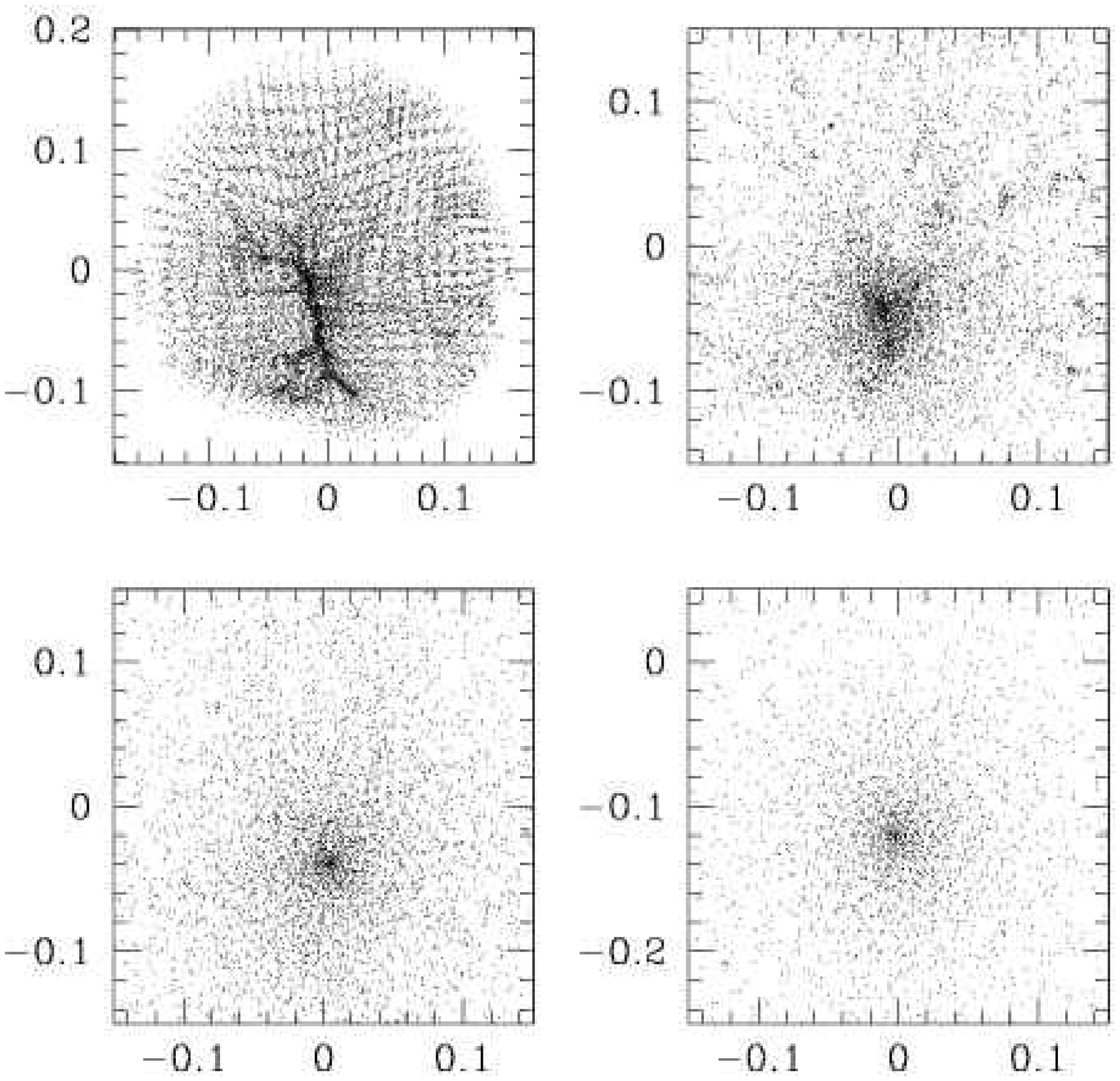}
\caption{The same as in Fig. \ref{pos1} but for the gas particles.}
\end{figure}

%%%%%%%%%%%%%%%%%%%%%%%%FIG 3
\begin{figure}
\centering
\includegraphics[width=7cm,height=7cm]{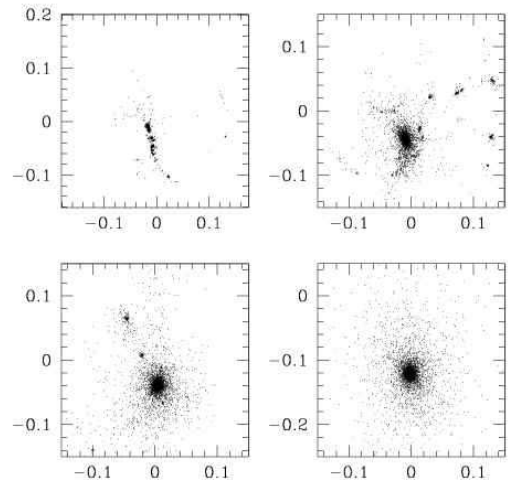}
\caption{The same as in Fig.\ref{pos1} but for the star particles.}
\end{figure}

\subsection{Chemical enrichment} \label{chim}

The material,  processed by nuclear reactions in stellar interiors
and given back to the interstellar medium by supernova explosions
and/or stellar winds, enriches the gas in metals. To correctly
estimate the chemical enrichment, first one has to carefully
calculate the stellar ejecta and the chemical yields per stellar
generation, second to suitably spread the newly formed metals into
the surrounding gas; see \citet{Lia02} for a complete description of
the physical processes involved in the computation and of their
implementation in the code (the laws governing the gas restitution
together with the chemical enrichment are treated in the same
probabilistic manner as the ones governing the star formation and
supernova rate processes).

The metals are distributed among the gas particles by means of the
\textit{diffusion equation}

\begin{equation}
\frac{dZ}{dt} = - \kappa \nabla^2 Z   \label{metal}
\end{equation}

\noindent where $\kappa$ is the diffusion coefficient, for which we
adopt the estimate by \citet{Lia02} $\rm \kappa = 9.25\times
10^{16}\,\, km^2 \,\,s^{-1}$ based on the  \citet{Thornton98}
models. In our simulations we explicitly follow the contribution to
the mass abundance of ten elements, namely Z (metals in general),
He, C, O, N, Mg, Si, S, Ca and Fe.

%%%%%%%%%%%%%%%%%%%%%%%%FIG 4
\begin{figure}
\centering
\includegraphics[width=7cm,height=7cm]{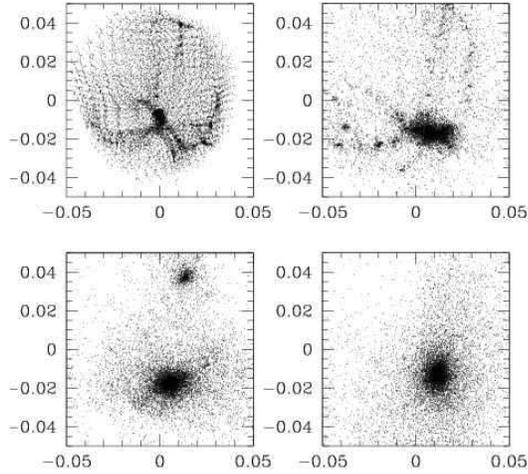}
\caption{Position on the $xy$ plane of the Dark Matter particles for
Model B (proper Mpc) at different times; from left to right and  top
to bottom, $z= 6.3, 4.2, 2.2, 1.0$.} \label{pos2}
\end{figure}

%%%%%%%%%%%%%%%%%%%%%%%%FIG 5
\begin{figure}
\centering
\includegraphics[width=7cm,height=7cm]{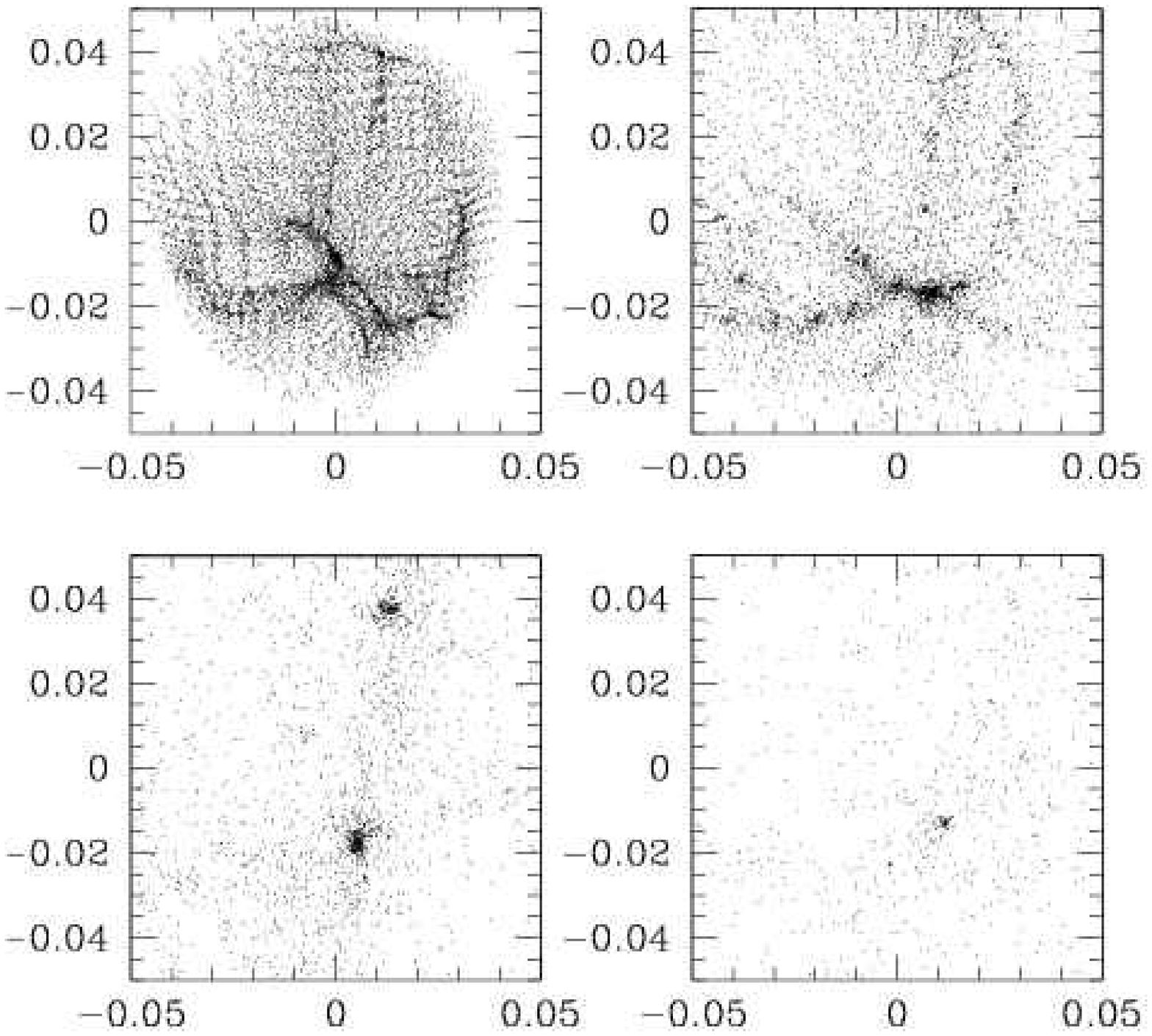}
\caption{The same as in Fig.\ref{pos2} but for the gas particles.}
\end{figure}

%%%%%%%%%%%%%%%%%%%%%%%%FIG 6
\begin{figure}
\centering
\includegraphics[width=7cm,height=7cm]{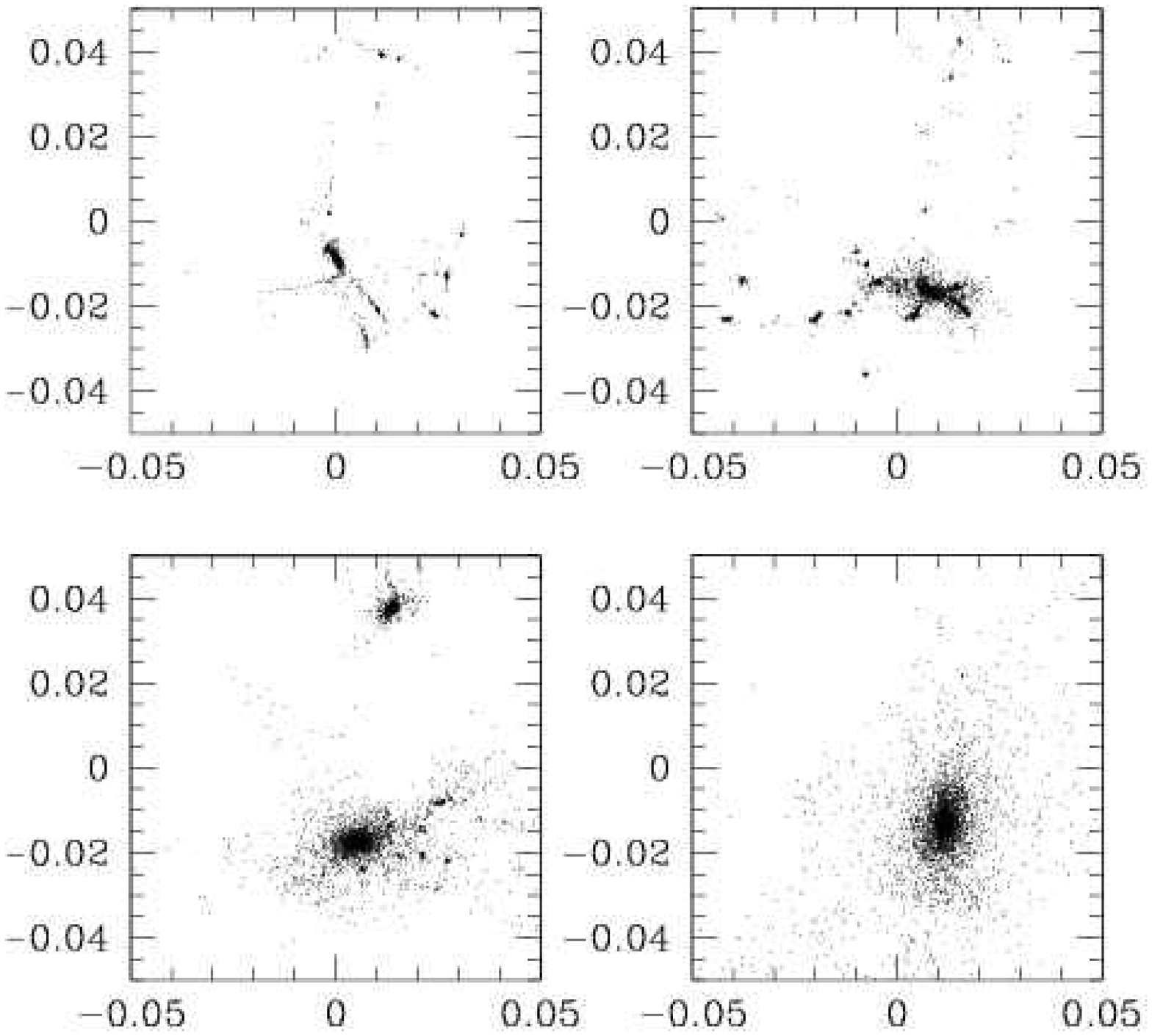}
\caption{The same as in Fig.\ref{pos2} but for the star particles.}
\label{posfin}
\end{figure}

%%%%%%%%%%%%%%%%%%%%%%%%FIG 7
\begin{figure}
\centering
\includegraphics[width=6.5cm,height=6.0cm]{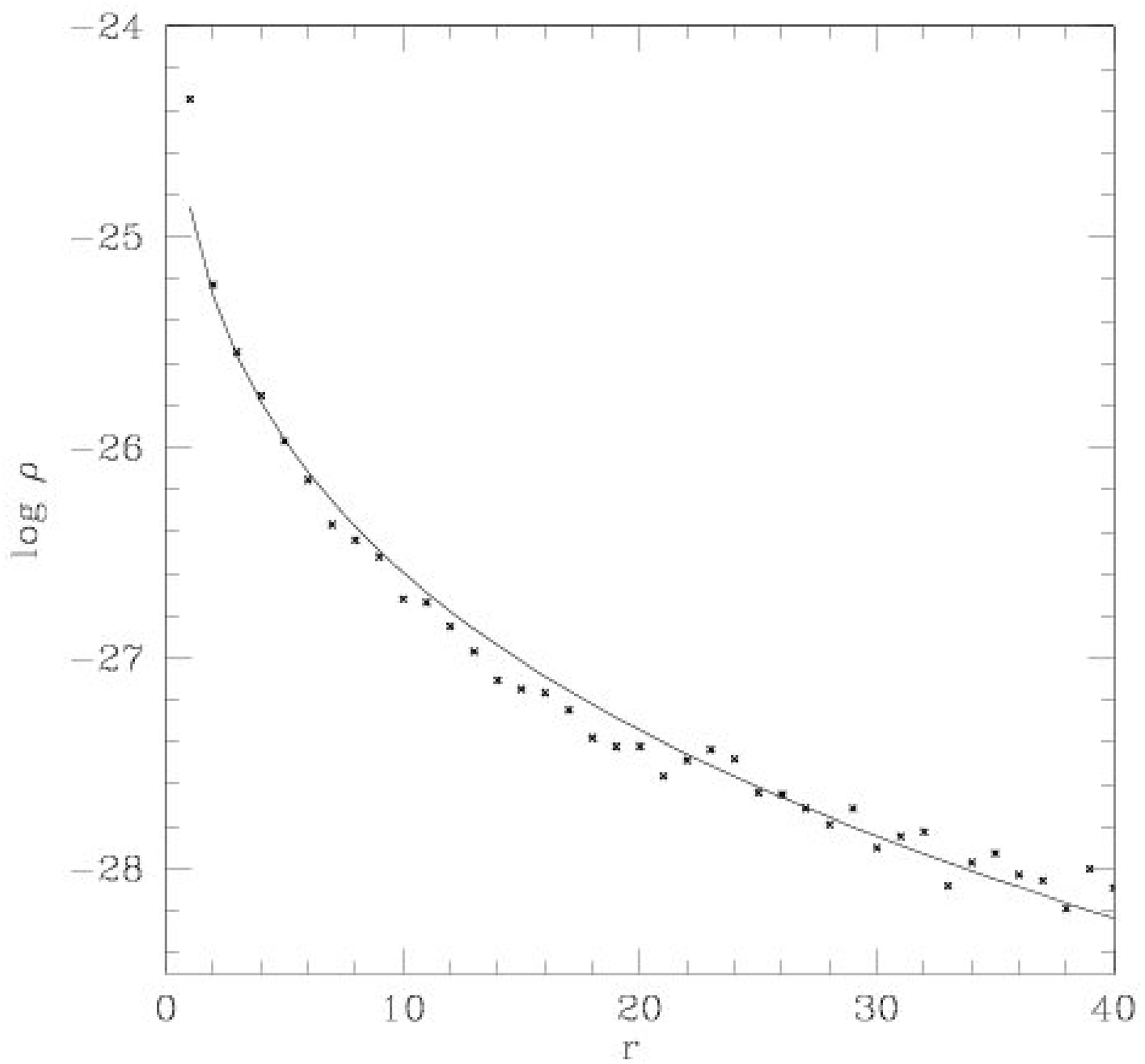}
\includegraphics[width=6.5cm,height=6.0cm]{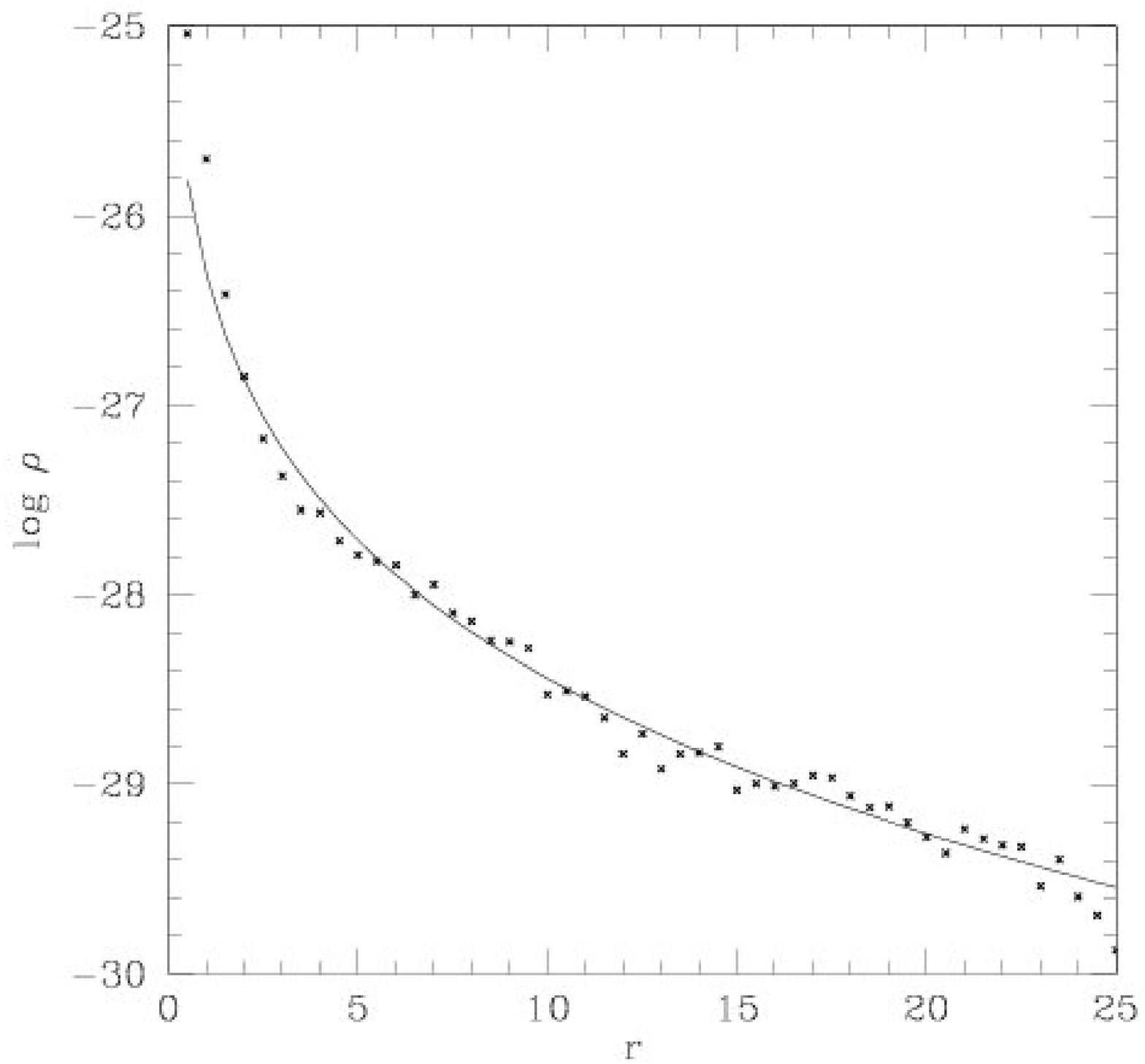}
\caption{ {\bf Top Panel}: Final surface density profile for  Model
A ($log[g/cm^2]$ vs kpc). The dots are averaged surface densities
projected on the $xy$ plane. The solid line is a Sersic law (see
text for details). {\bf Bottom Panel}: The same but for Model B. }
\label{vaucA_vaucC}
\end{figure}

%%%%%%%%%%%%%%%%%%%%%%%%FIG 8
\begin{figure}
\centering
\includegraphics[width=6.5cm,height=6.0cm]{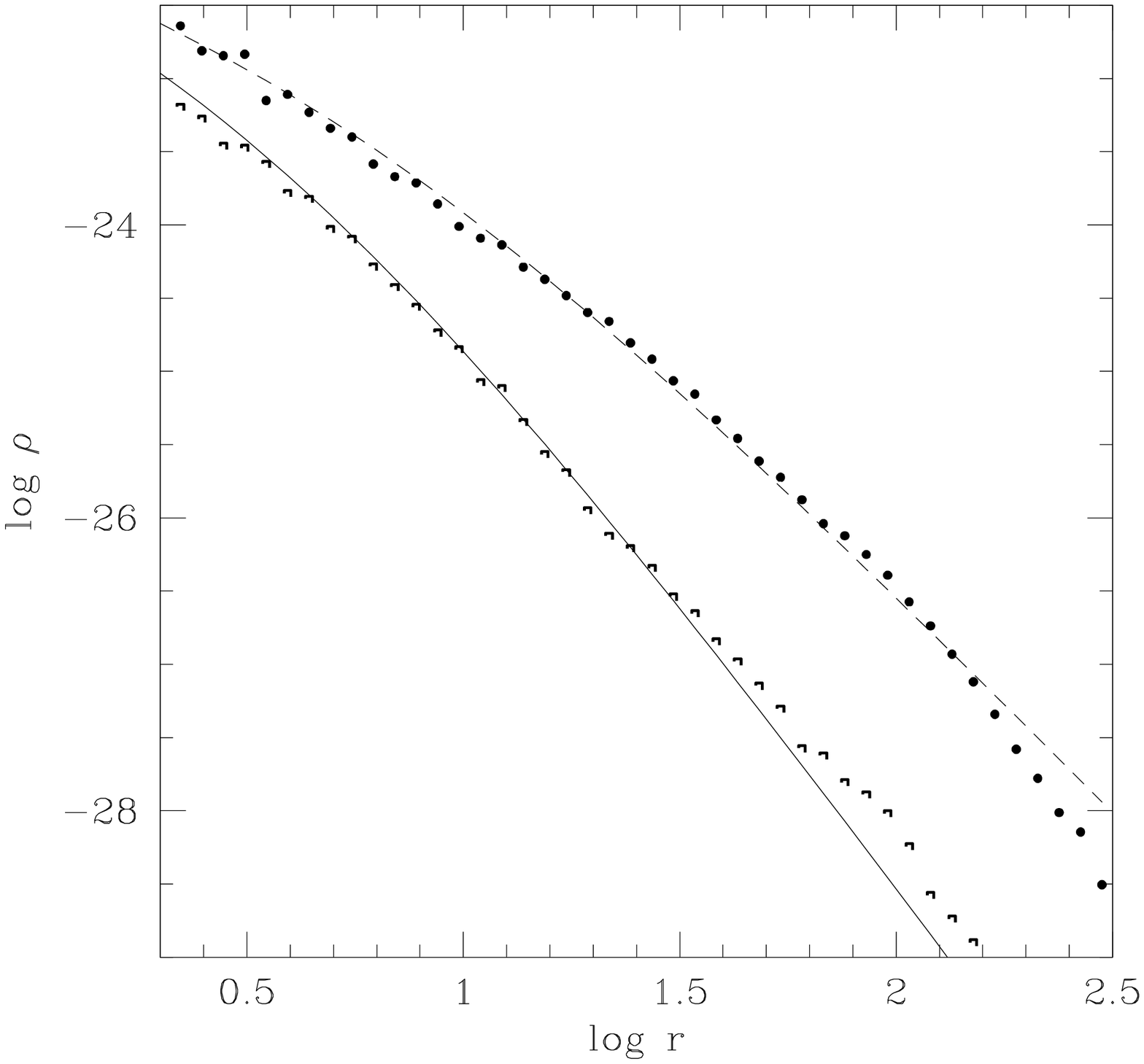}
\includegraphics[width=6.5cm,height=6.0cm]{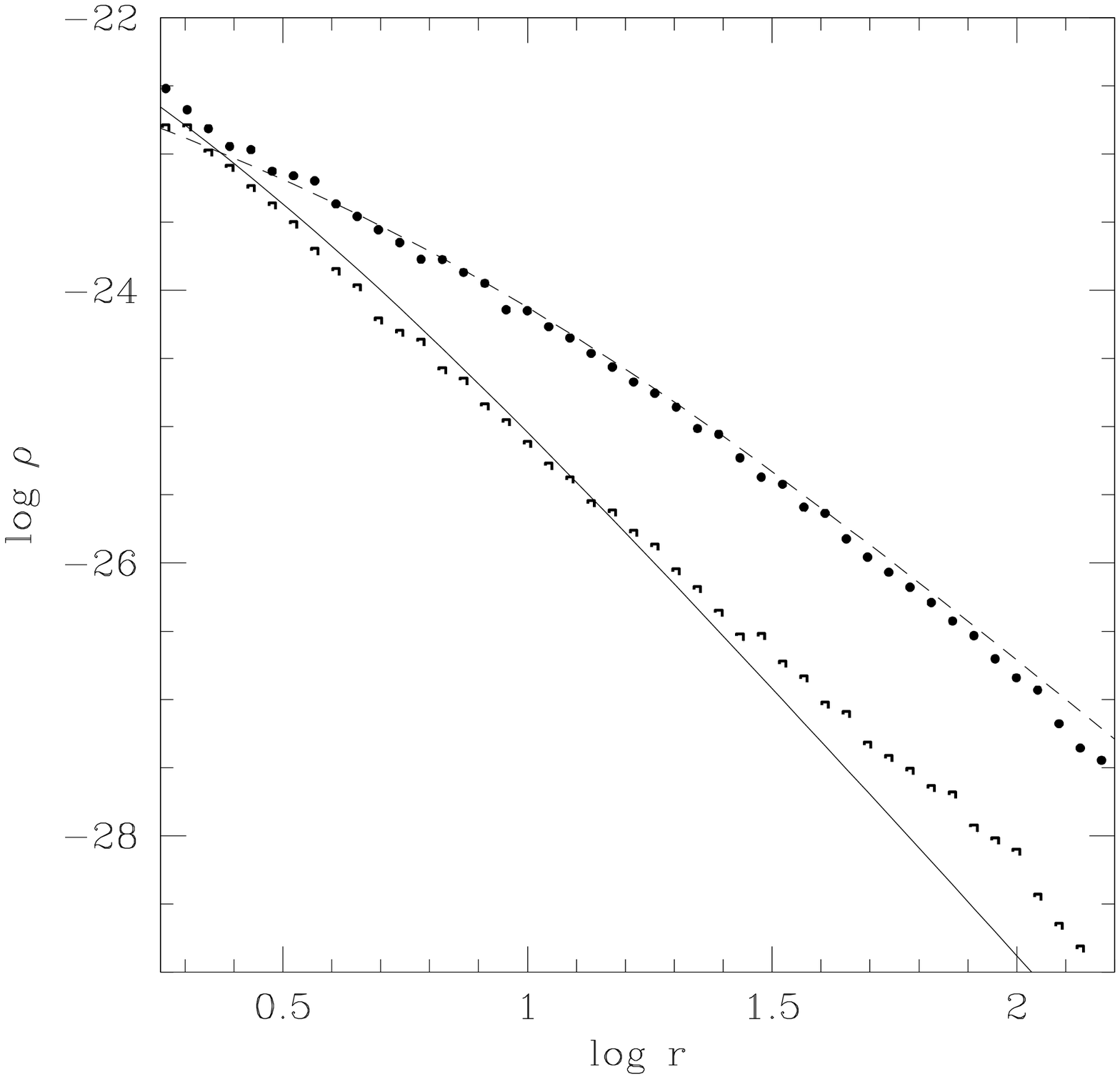}
\caption{ {\bf Top Panel}: Final spherical density profile for Model
A ($log[g/cm^3]$ vs $log[kpc]$). Dots and squares are respectively
CDM and stars averaged densities in spherical shells at the
corresponding radius. The solid line is the Hernquist law; the
dotted line is the NFW profile. {\bf Bottom Panel}: The same but for
Model B. } \label{densA_densC}
\end{figure}

%%%%%%%%%%%%%%%%%%%%%%%%FIG 9
\begin{figure}
\centering
\includegraphics[width=6.5cm,height=6.0cm]{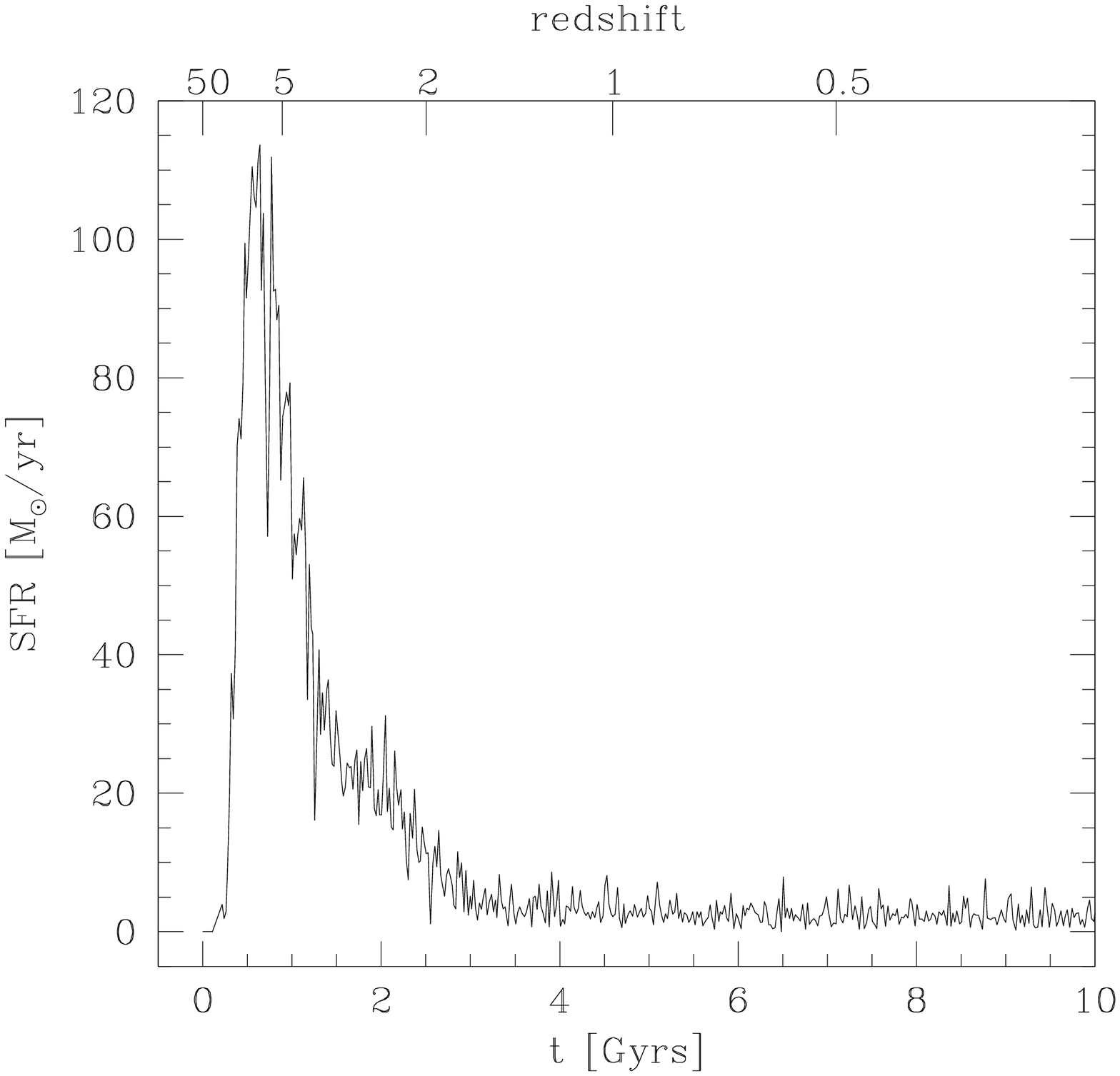}
\includegraphics[width=6.5cm,height=6.0cm]{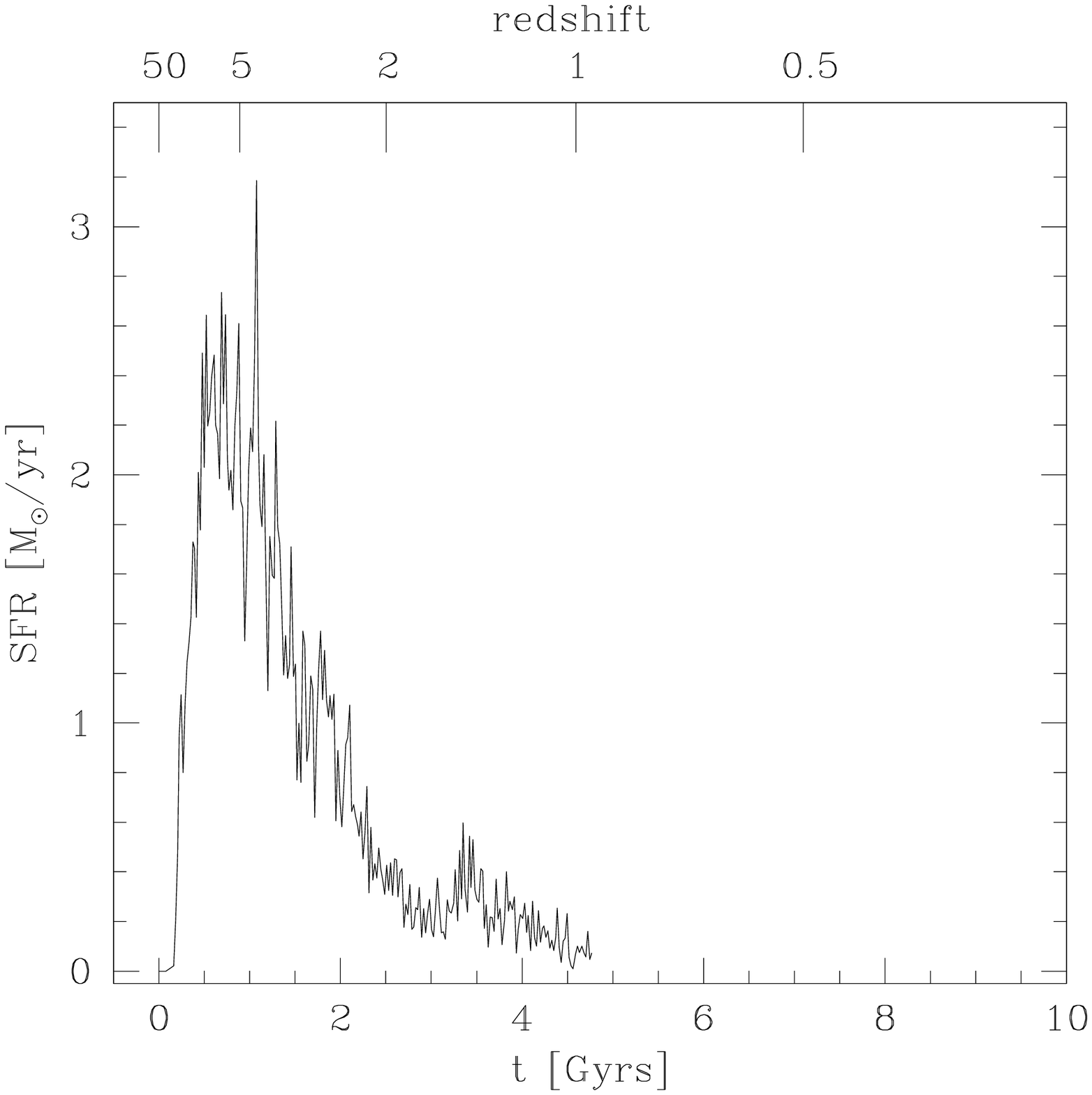}
\caption{{\bf Top Panel}: Star formation rate ($M_\odot$ per year)
in Model A. {\bf Bottom Panel}: The same but for Model B. }
\label{sfrA_sfrC}
\end{figure}

%%%%%%%%%%%%%%%%%%%%%%%%FIG 10
\begin{figure}
\centering
\includegraphics[width=6.5cm,height=6.0cm]{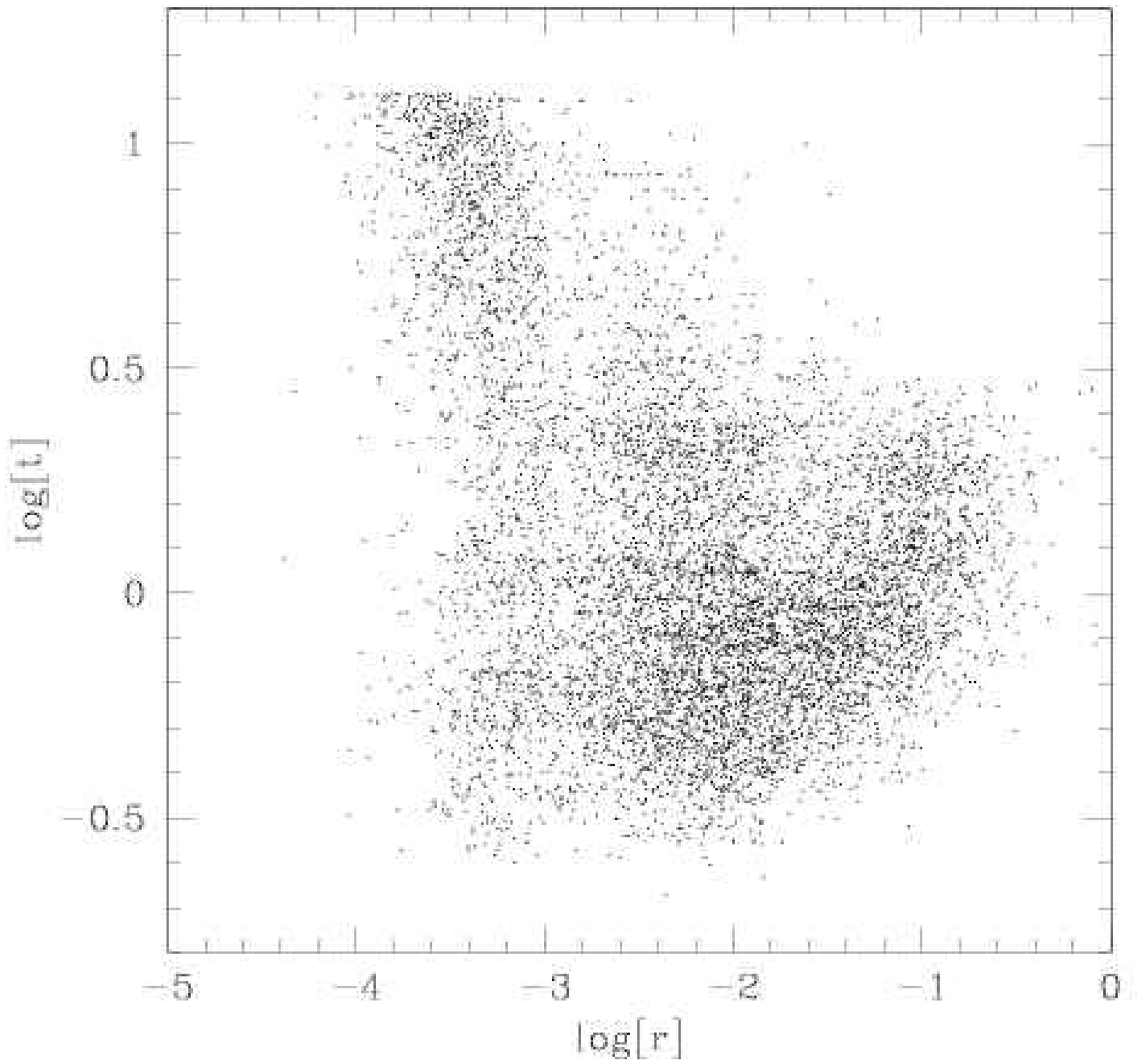}
\includegraphics[width=6.5cm,height=6.0cm]{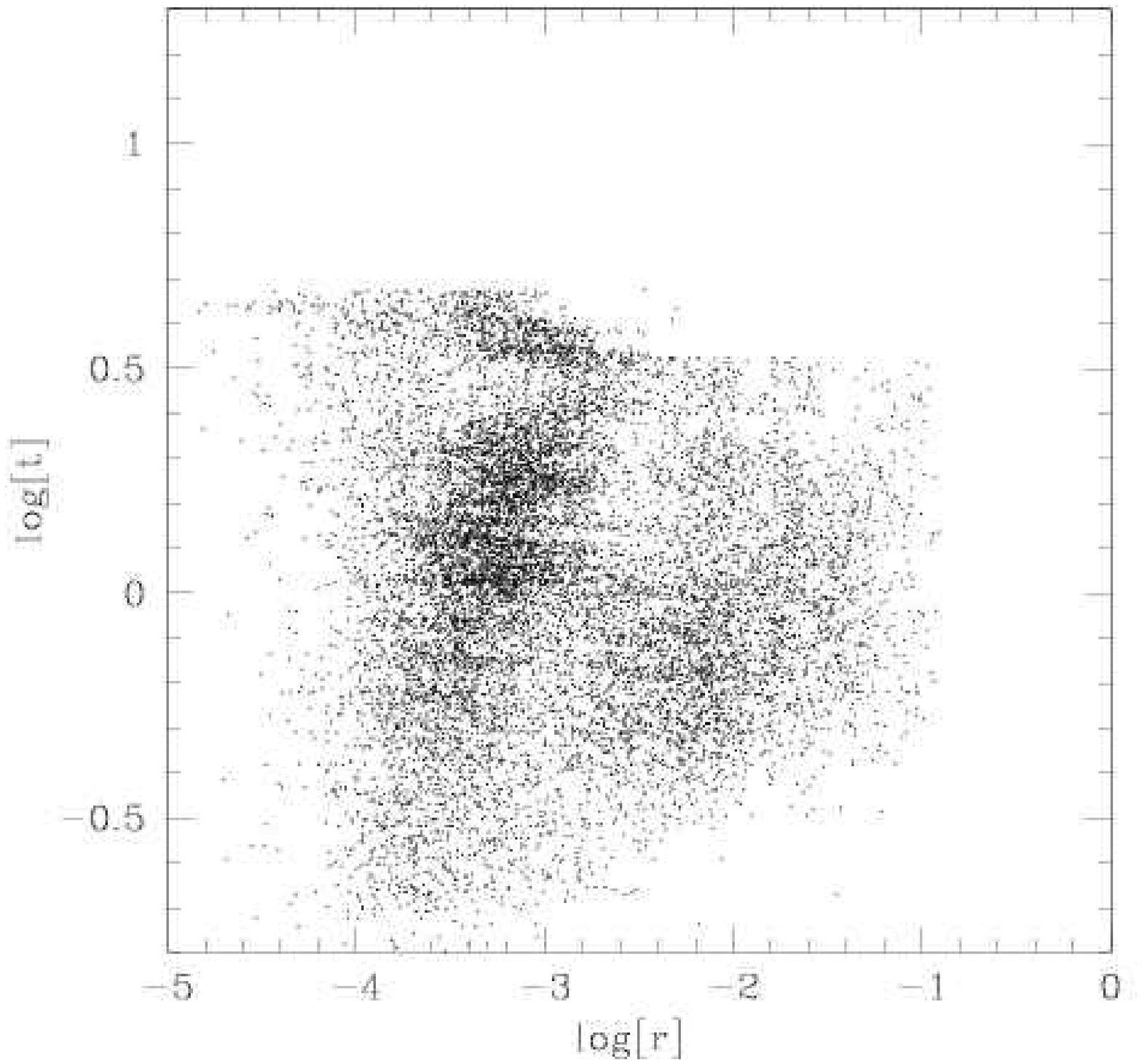}
\caption{ {\bf Top Panel}: Formation time (log[Gyr]) vs radial
distance (log[Mpc]) of the stars in Model A. It is easy to notice
how the galaxy we see today is gradually built up moving from the
external regions toward the  centre where star formation lasts very
long. {\bf Bottom Panel}: the same but for Model B.}
\label{ageradA_ageradC}
\end{figure}

%%%%%%%%%%%%%%%%%%%%%%%%FIG 11
\begin{figure}
\centering
\includegraphics[width=6.5cm,height=6.0cm]{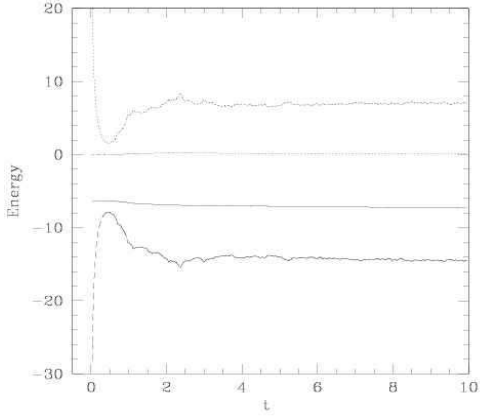}
\caption{Variation of  (from top to bottom) kinetic, thermal, total
and potential  energy in code units as function of time in Gyrs for
Model A (all components).} \label{eneA_eneB}
\end{figure}

%%%%%%%%%%%%%%%%%%%%%%%%FIG 12
\begin{figure}
\centering
\includegraphics[width=6.5cm,height=6.0cm]{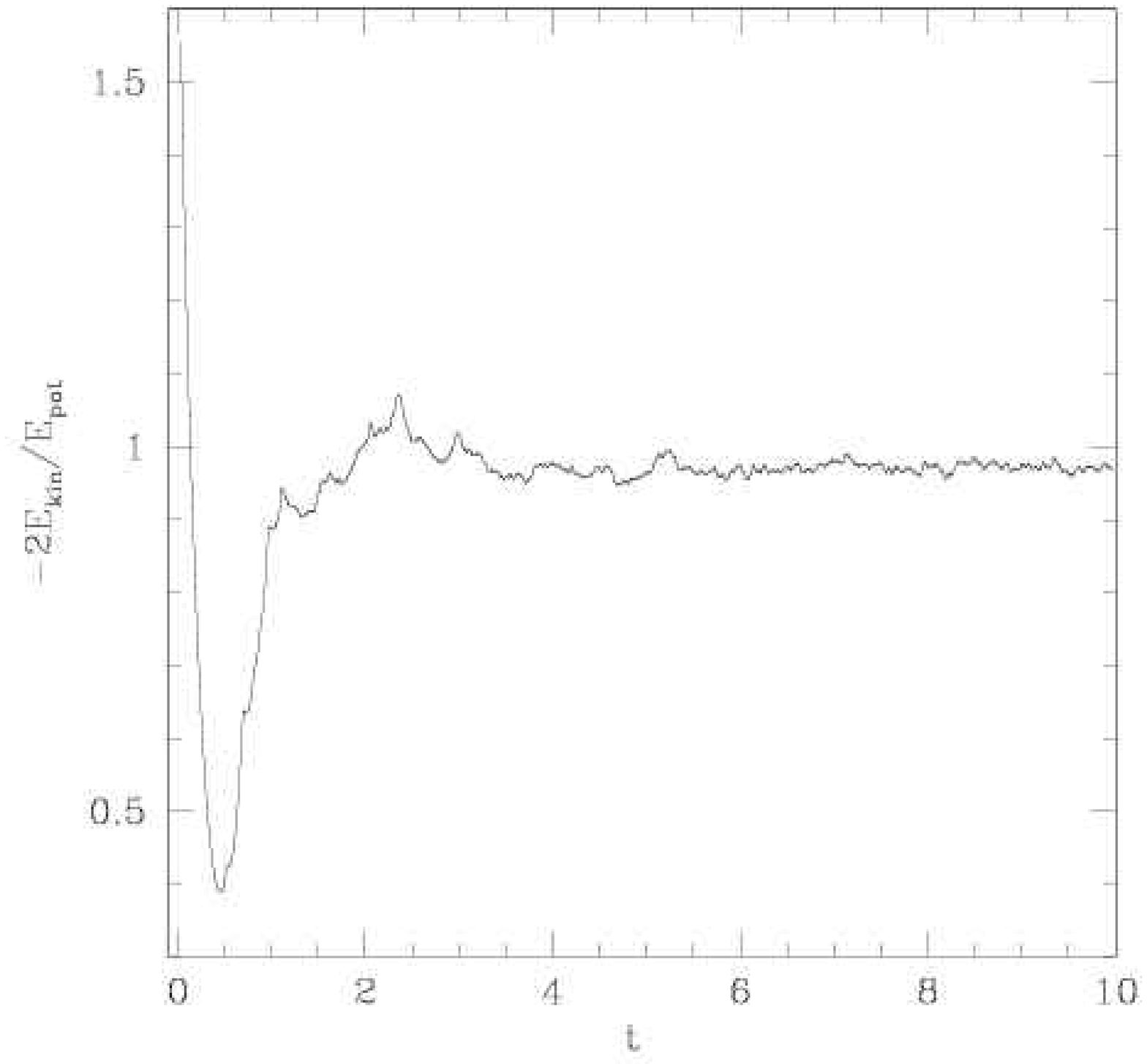}
\includegraphics[width=6.5cm,height=6.0cm]{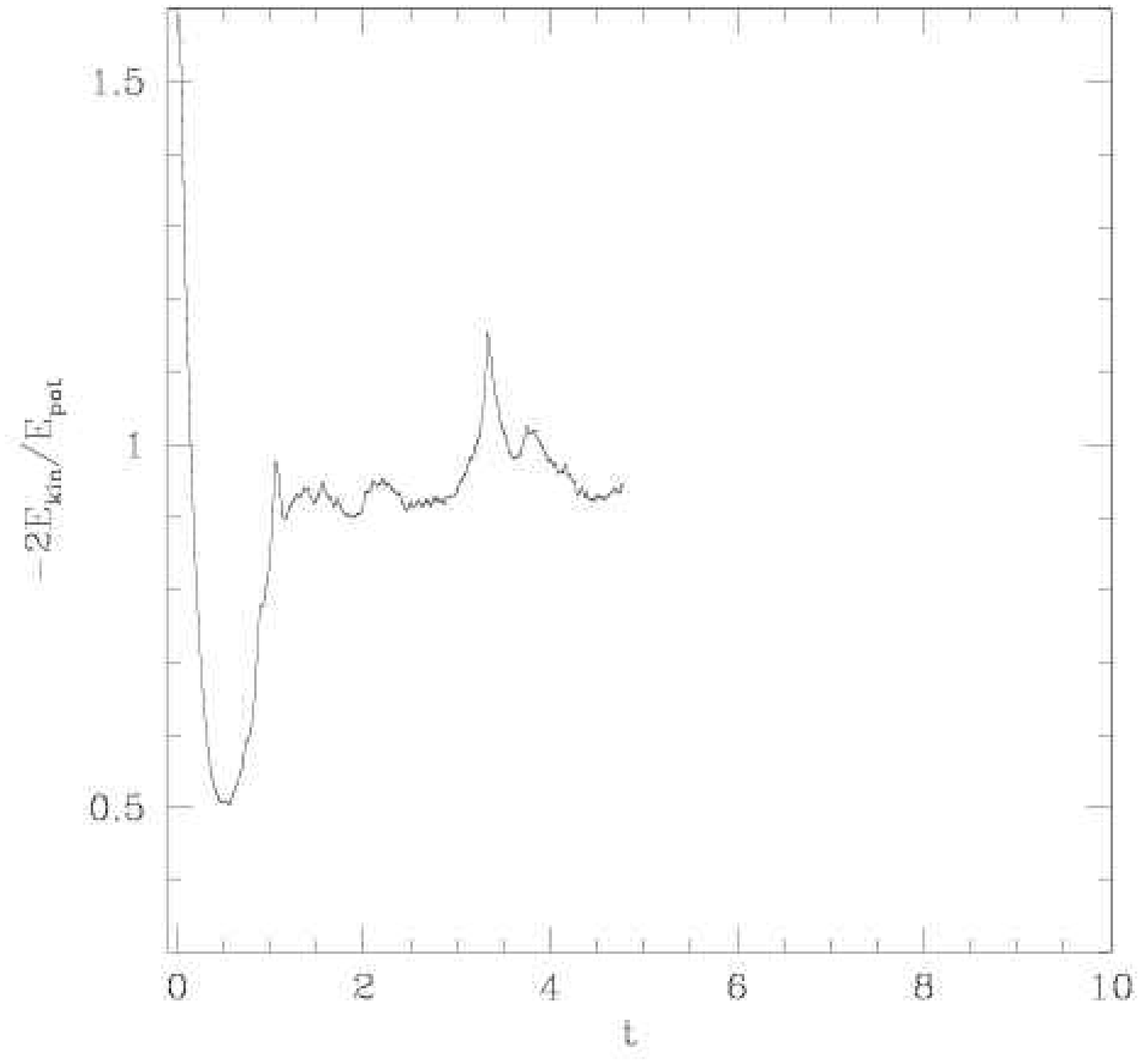}
\caption{{\bf Top Panel}: Trend of the Virial trace
$-2E_{kin}/E_{pot}$ for Models A and B from top to bottom. Ages are
in Gyr.} \label{virA_virC}
\end{figure}

%%%%%%%%%%%%%%%%%%%%%%%%FIG 13
\begin{figure}
\centering
\includegraphics[width=6.5cm,height=6.0cm]{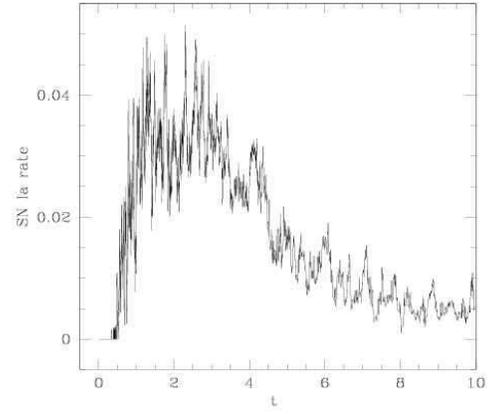}
\includegraphics[width=6.5cm,height=6.0cm]{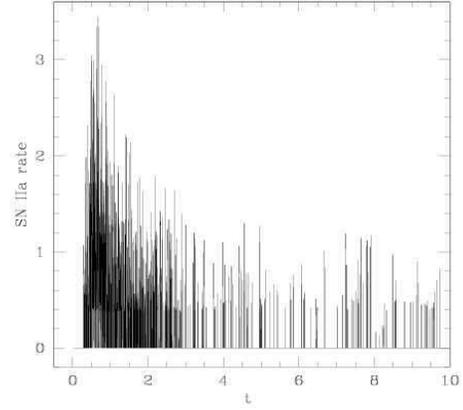}
\caption{ {\bf Top Panel } {\bf Type Ia supernova rate (explosions
per year)} as function of time for Model A. {\bf Bottom panel}: the
same as above but for Type II supernovae.} \label{snIA_snIIA}
\end{figure}

%%%%%%%%%%%%%%%%%%%%%%%%FIG 14
\begin{figure}
\centering
\includegraphics[width=6.5cm,height=6.0cm]{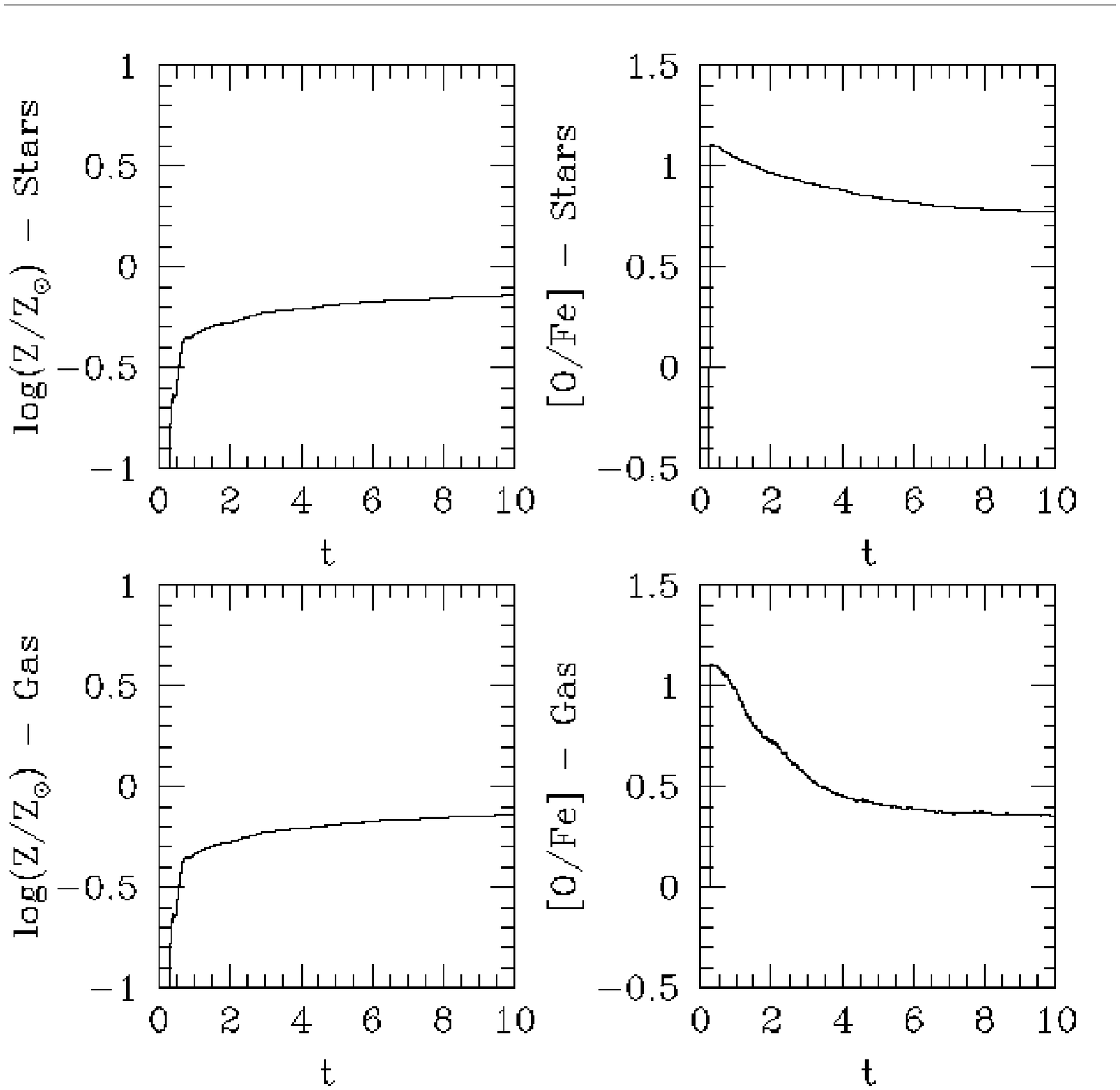}
\includegraphics[width=6.5cm,height=6.0cm]{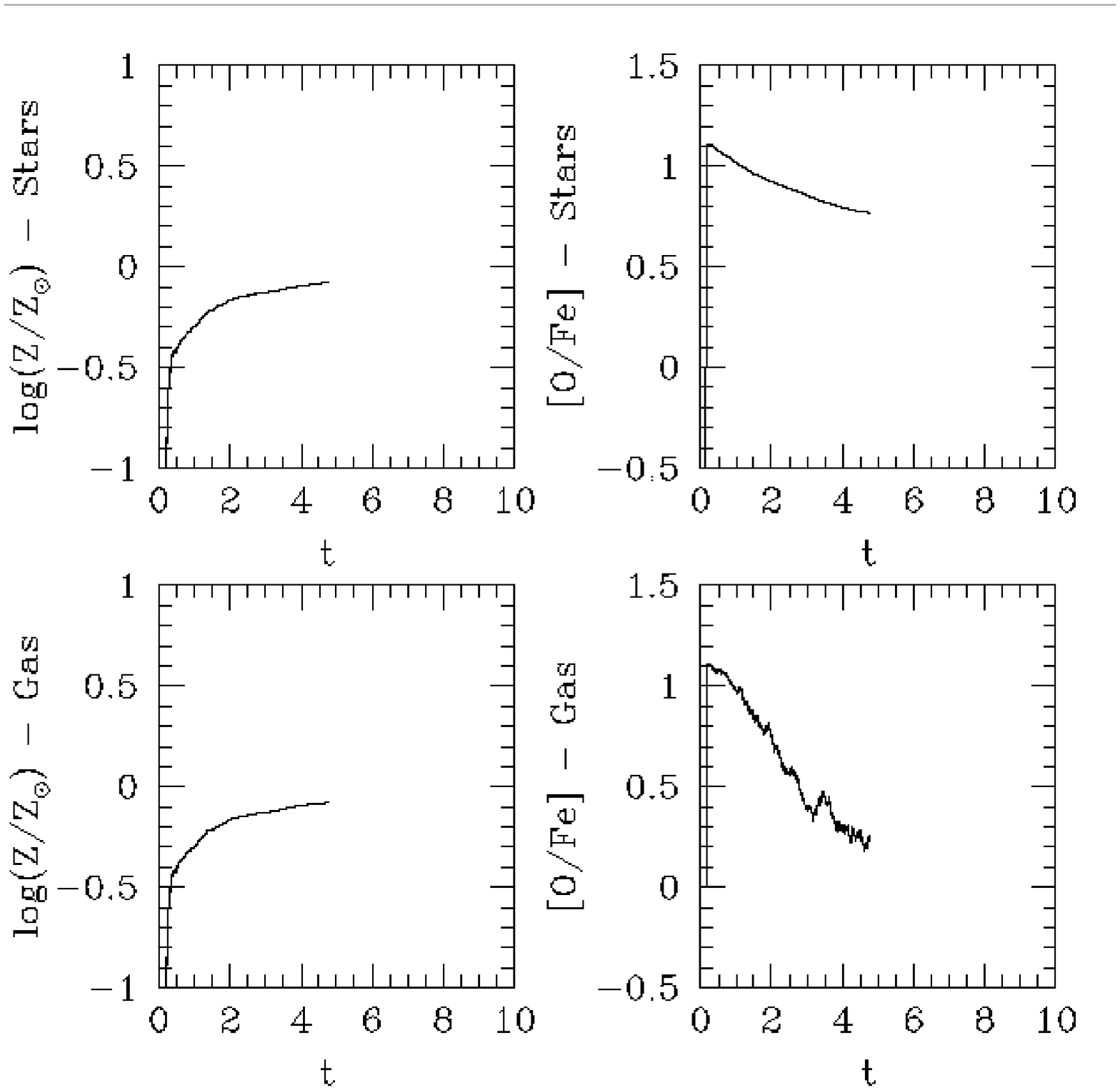}
\caption{ {\bf Top Panel}: Chemical evolution of the star and gas
particles in  Model A as a function of the age;  left panels show the
mean total metallicity Z in solar units, right panels the ratio [O/Fe].
The age is in Gyrs. {\bf Bottom Panel}: The same as above but for Model B. }
\label{metA_metC}
\end{figure}

%%%%%%%%%%%%%%%%%%%%%%%%FIG 15
\begin{figure}
\centering
\includegraphics[width=6.5cm,height=6.0cm]{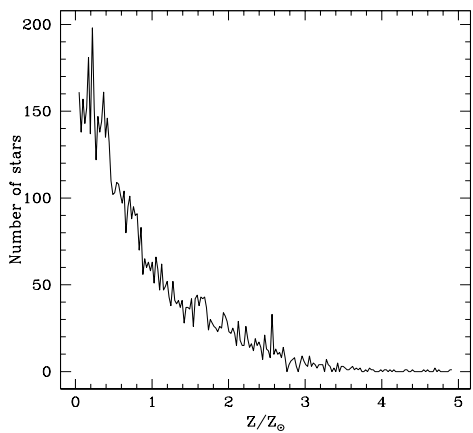}
\caption{Number distribution of stars in metallicity at the end of the simulation for Model A.}
\label{chimbinA}
\end{figure}

%%%%%%%%%%%%%%%%%%%%%%%%%%FIG 16
\begin{figure}
\centering
\includegraphics[width=6.5cm,height=6.0cm]{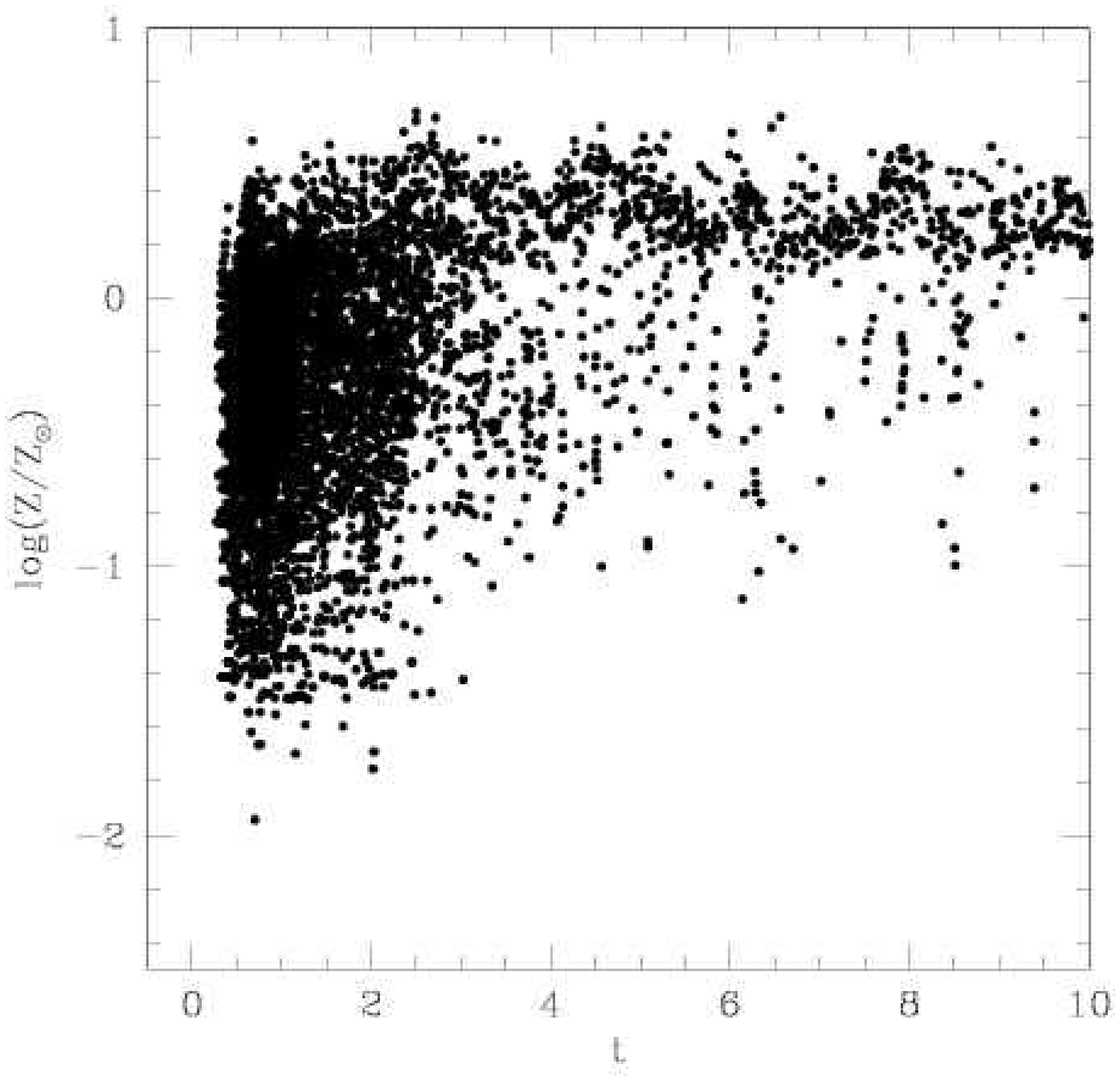}
\includegraphics[width=6.5cm,height=6.0cm]{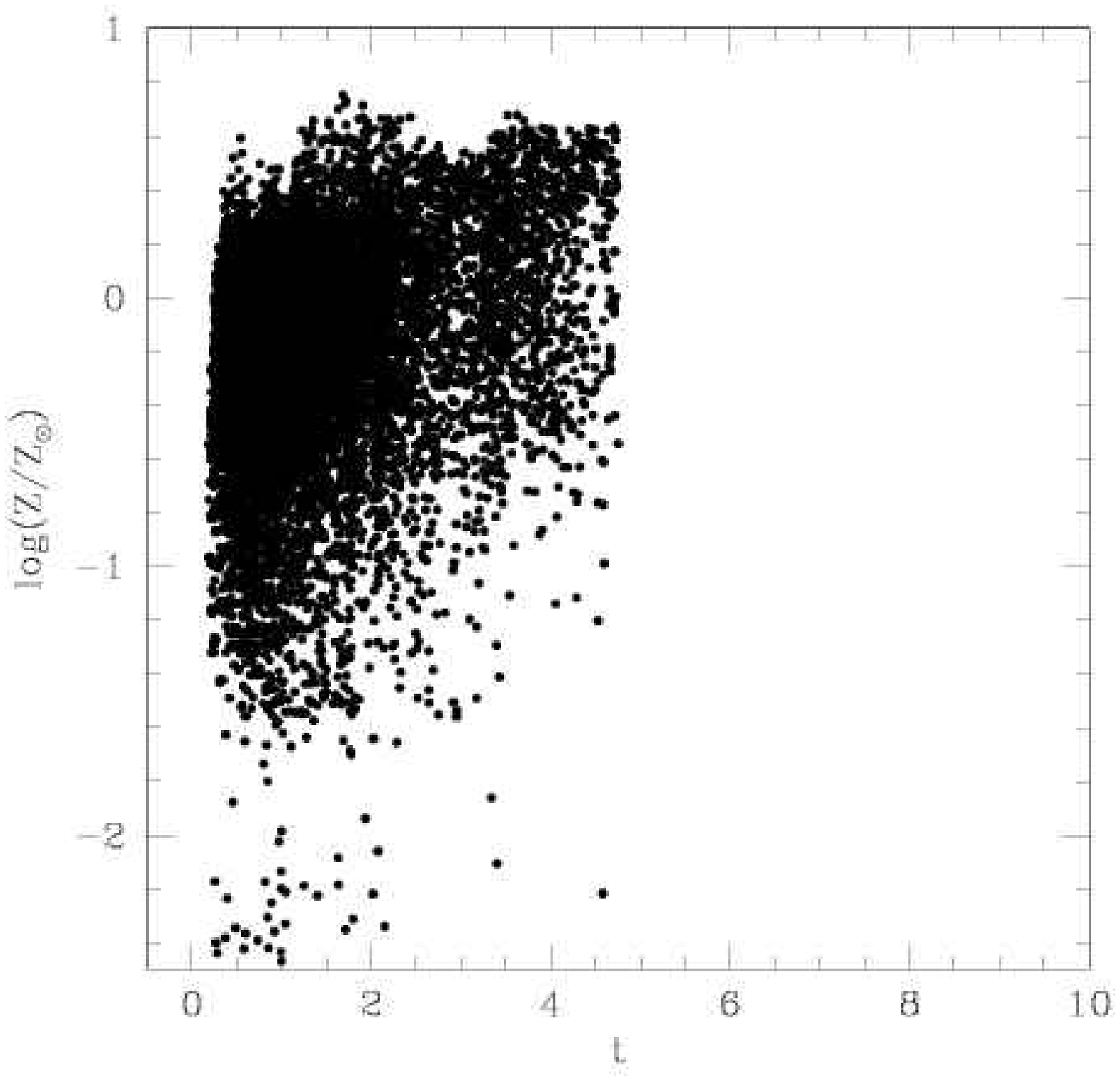}
\caption{ {\bf Top Panel}: Metallicity ($log[Z/Z_\odot]$) vs age of
formation (Gyrs) of the stars in Model A. {\bf Bottom Panel}: The
same but for Model B.} \label{timechimA_timechimC}
\end{figure}

%%%%%%%%%%%%%%%%%%%%%%%%FIG 17
\begin{figure}
\centering
\includegraphics[width=6.5cm,height=6.0cm]{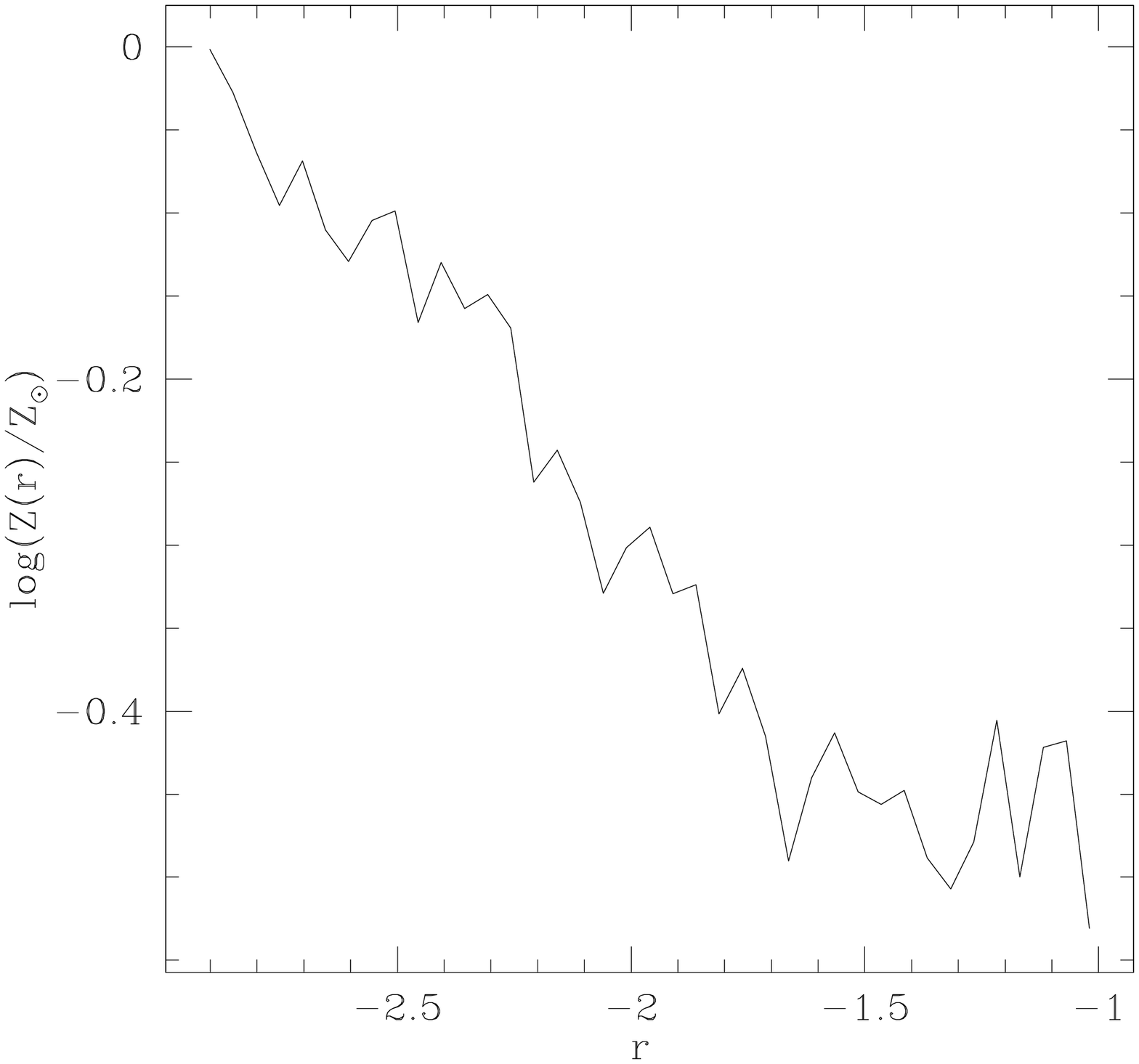}
\includegraphics[width=6.5cm,height=6.0cm]{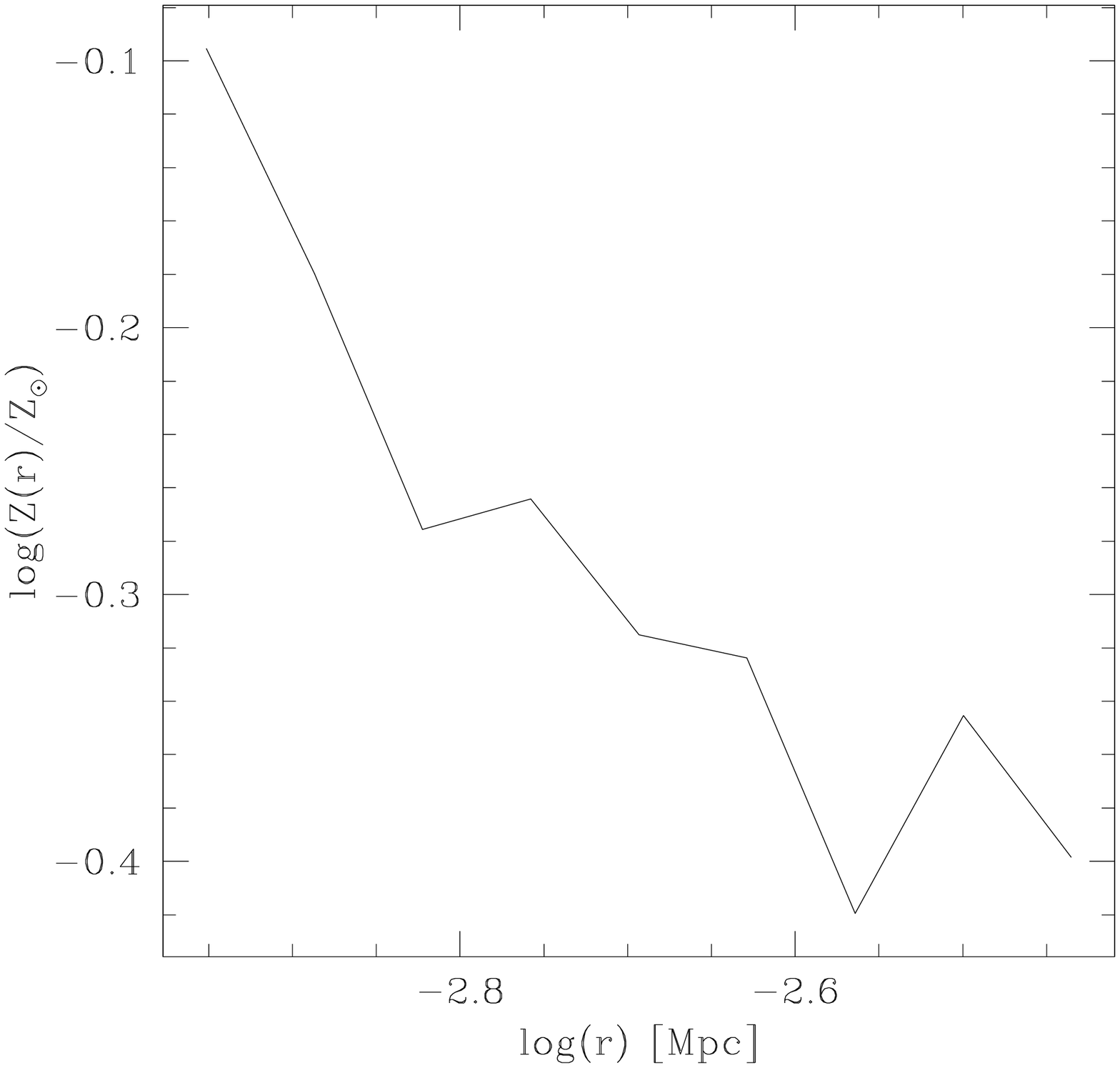}
\caption{ Radial gradient in the spherically averaged metallicity
for the stars of Model A (top) and B (bottom). Radii are in
$log[Mpc]$).} \label{radmetA}
\end{figure}

%%%%%%%%%%%%%%%%%%%%%%%%FIG 18
\begin{figure}
\centering
\includegraphics[width=6.5cm,height=6.0cm]{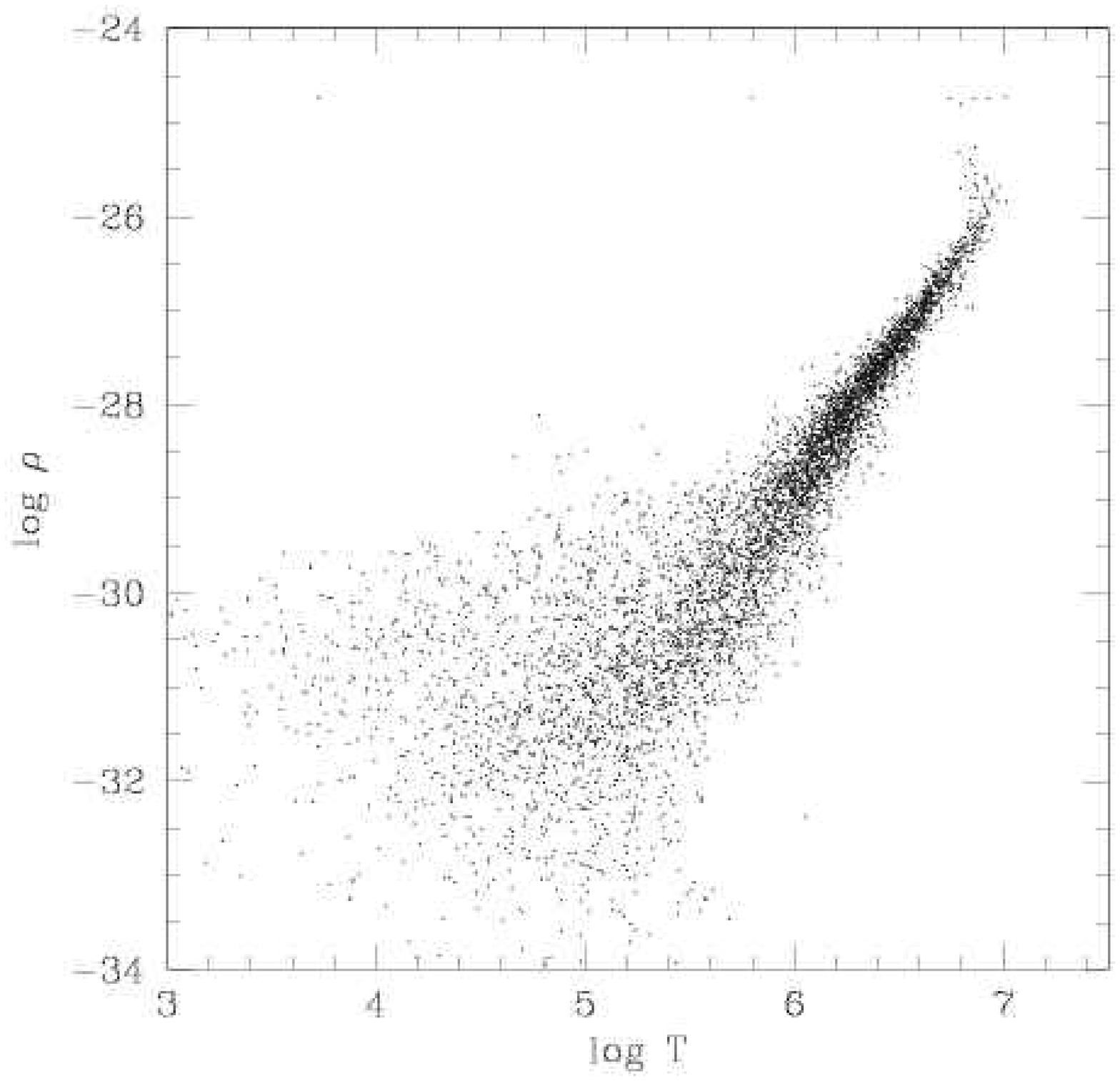}
\includegraphics[width=6.5cm,height=6.0cm]{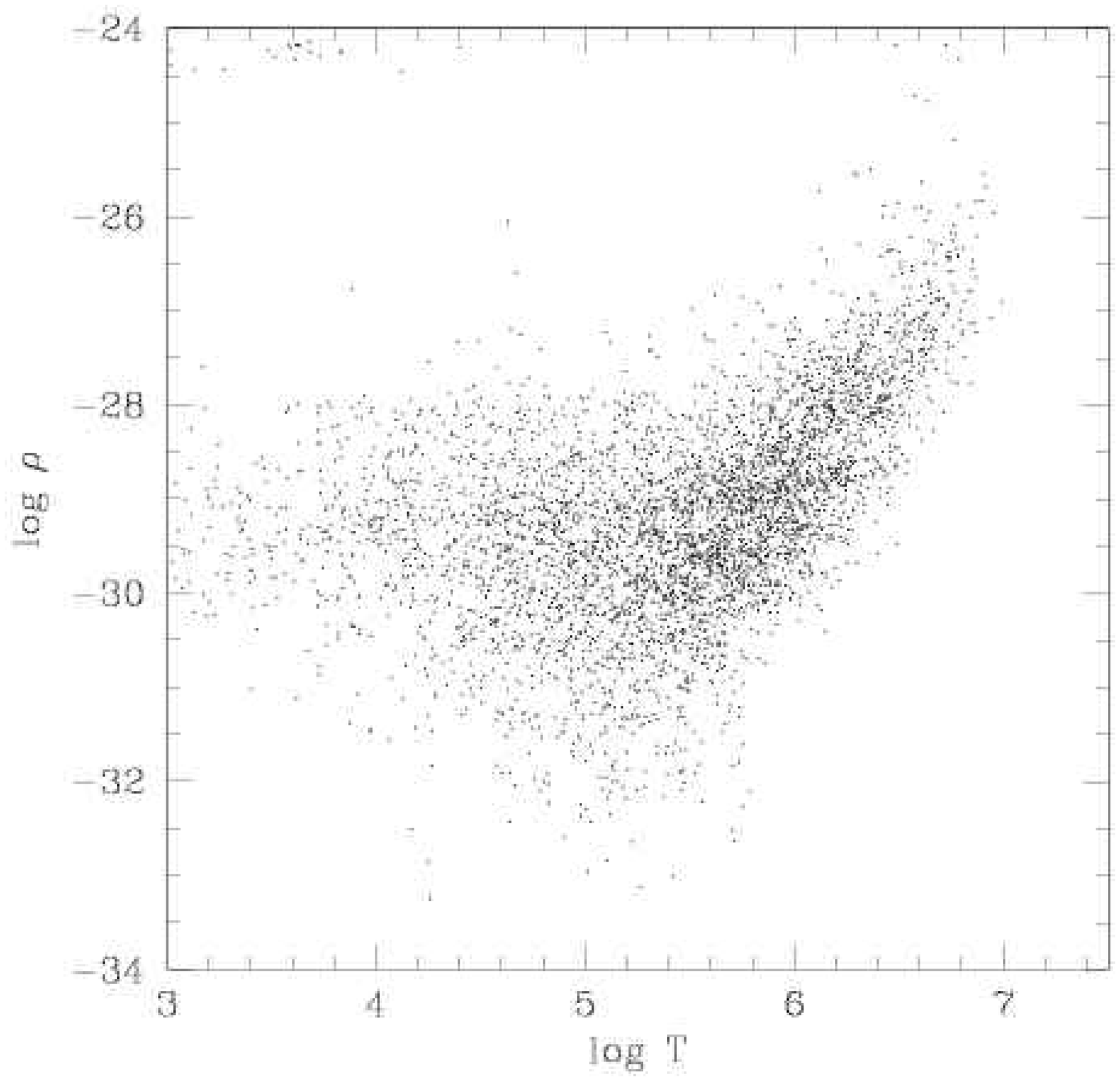}
\caption{ {\bf Top Panel}:  Gas density ($log[g/cm^3]$) vs
temperature ($log[K]$) of Model A. {\bf Mid Panel}: The same but for
Model B. }
\label{tempdensA_tempdensC}
\end{figure}

%%%%%%%%%%%%%%%%%%%%%%%%%%FIG 19
\begin{figure}
\centering
\includegraphics[width=6.5cm,height=6.0cm]{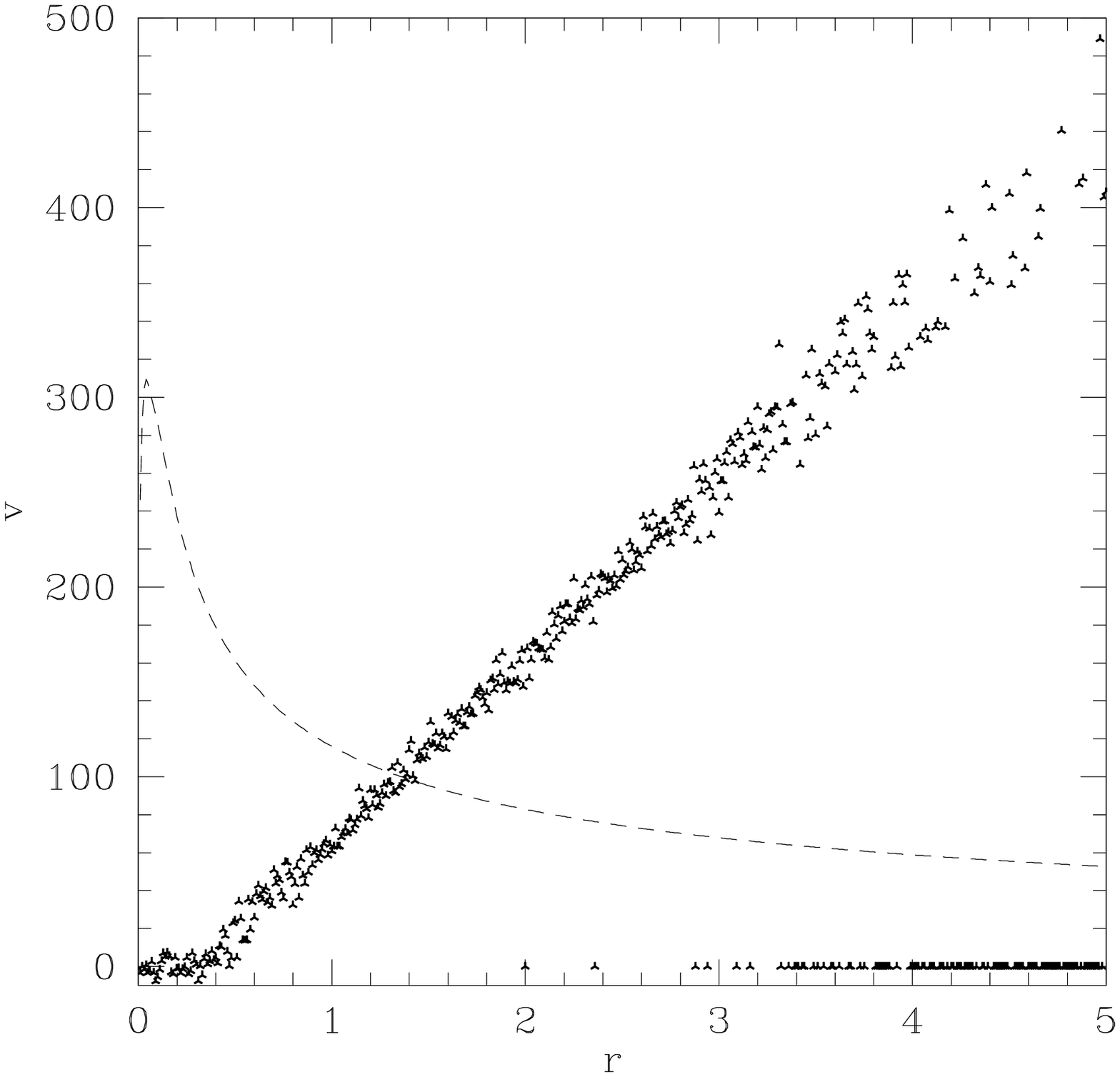}
\includegraphics[width=6.5cm,height=6.0cm]{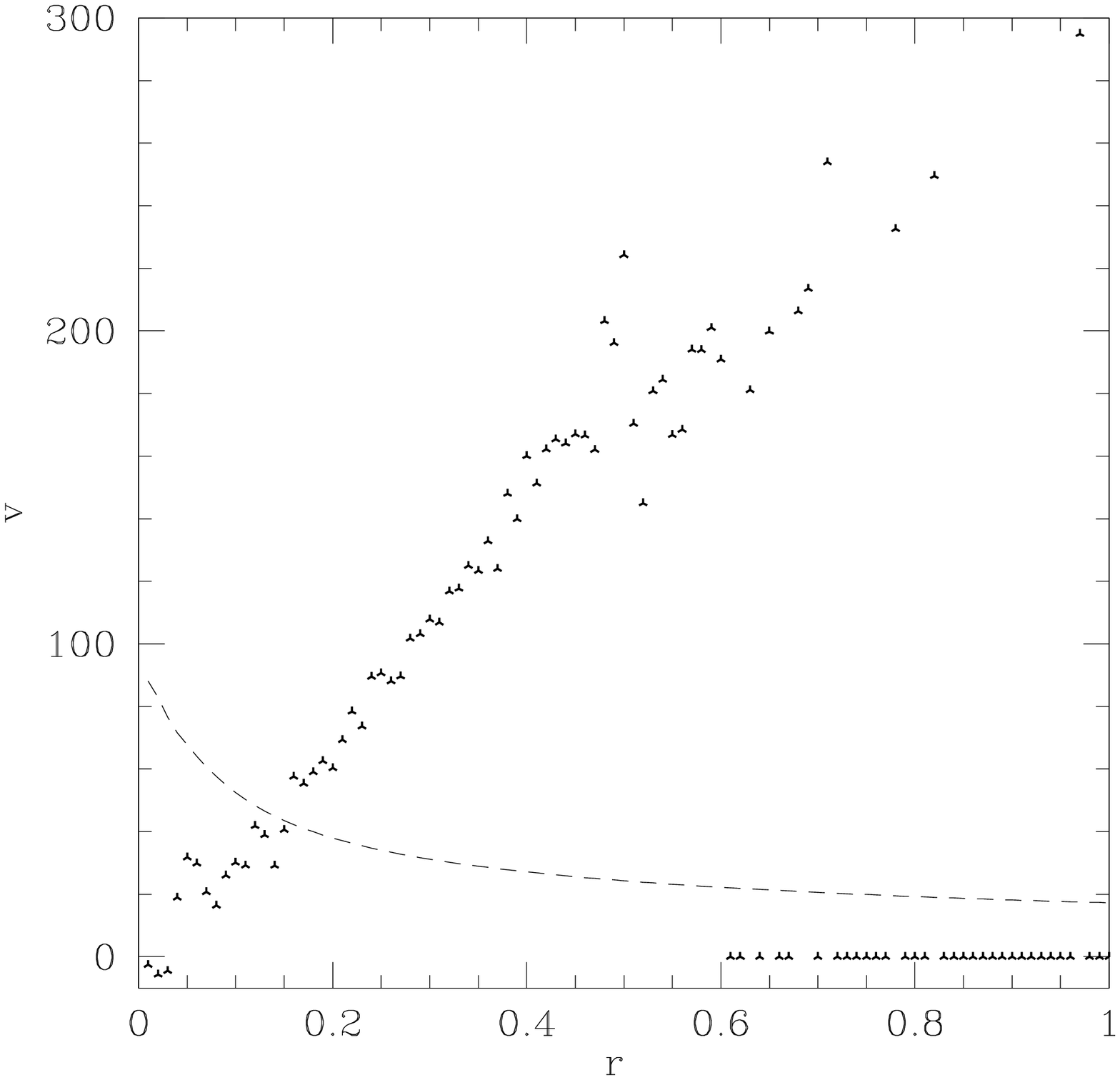}
\caption{ {\bf Top panel}: Radial velocity (km/sec, points) and escape
velocity (dotted line), vs radius (Mpc) for  Model A. Both
velocities are in proper km/s. {\bf Bottom panel}: The same but for
model B. }
\label{winds}
\end{figure}

\section{Galaxy models: the parameters}\label{two_models}

\subsection{The cosmological initial conditions}

As already pointed out, in this paper we present two simulations,
whose parameters are summarized in Table \ref{tabcosmo}.

All simulations are framed in a \textit{flat Universe} with spectral
index for the initial perturbations $n$=1. Denoting with $\Omega_i$
the ratio between density of the generic component $i$ and the
critical density to close the Universe, in our Standard CDM models
we adopt  $\Omega_b = 0.1$ (for baryons) and $\Omega_c = 0.9$ (for
Dark Matter). The Hubble constant is taken to be $\rm 50\, km\,
sec^{-1}\, Mpc^{-1}$.

Little is said in literature about the choice of the initial
temperature to be assigned to baryonic particles. In general, the
choice depends on the temperature coverage of the cooling processes
under consideration, and very often the initial temperature is set
at the lowest value of the temperature interval.  This is less of a
problem with our code because we have implemented the large
tabulations supplied by \citet{Chiosi98} which extend down to
molecular processes at low temperatures (as low as about 10 -- 100
K).

A rough estimate of the mean background temperature of the Universe
at the redshift we are interested is derived as follows. At the
\textit{decoupling} redshift $z \sim 1000$, the temperature is about
3000 K. After  decoupling, the Thomson scattering ceases, but a
sufficiently large fraction of free electrons remains to keep matter
in equilibrium with radiation via Compton scattering until $z \sim
160$ \citep{Madau03}. Since then the temperature ever decreases (as
far as   matter component are concerned) proportionally to
$(1+z)^2$. Given that we are considering over-dense regions, we
assume the initial temperature of the sphere to be equal to the mean
temperature of the Universe, at the time in which the mean
background density was equal to the mean density of our
proto-galactic sphere. With the initial red-shifts we have adopted
(in the range $50 < z < 100$), the temperature falls in the interval
10 to 150 K.

\subsection{Final parameters for the galaxy models}

For all simulations the initial \textit{smoothing length} parameter
for the baryonic component has been set equal to ten times the
\textit{softening length}. The \textit{softening length} $\epsilon$
in turn has been chosen in such away that both a good resolution and
the  \citet{Evrard88} relation are secured. This latter requires

\begin{equation}
\frac{Gm}{\epsilon} \ll \frac{GM}{R}
\end{equation}

\noindent i.e. $\epsilon \gg R/N_{part}$. In Table \ref{tabdyn} are
listed the dynamical and computational parameter  we have adopted
for the three models (the cosmological parameters have already been
given in Table  \ref{tabcosmo}).

The basic units for length, time ad mass used to express all
physical quantities entering the models are:

\begin{itemize}
\item  length: 1 proper Mpc;
\item  time: the lifetime of the Universe, corresponding to the cosmological model
adopted for the simulation;
 \item  mass: the unitary mass is
obtained supposing the free-fall time-scale of a sphere with unitary
radius equal to the unit time so that the gravitational constant
$G=1$ in code units.
\end{itemize}

\section{Galaxy models: the results} \label{results}

Model A has been evolved up to age of about 13 Gyr  (i.e. the age of
the Universe), whereas Model B  has been followed up to 5  Gyr.
Nevertheless,  the last computed model is well beyond the stage of
maximum stellar activity, i.e. the galaxies have already settled
down to a passive evolutionary regime.

\subsection{Spatial evolution}

The series of Figs. \ref{pos1} through \ref{posfin} show the
positions of Dark Matter, gas and star particles projected on the
$xy$ plane at different values of the redshift for Models A and B as
indicated. It must be noticed that the panels for Model A  has
length scale of 300 kpc along both axes, whereas the length is 100
kpc in the case of Model B.

One may sketch the following evolutionary picture for  the two
galaxies. The proto-galaxy initially expands with the Hubble flow
and the expansion becomes slower and slower until it detaches from
it first in the central regions (or in general in the denser
regions) and later in the outskirts. Not all gas is able to collapse
but a certain (often huge) fraction remains at the edges of the just
born galaxy. This is the result of the inhomogeneous distribution of
the initial densities that lead some regions to collapse earlier
than others independently of their distance from the centre of the
system. As a whole, the collapse deviates from the ideal spherical
case. In particular, one may stress that the final baryonic-to-Dark
ratio will be different from the initial one, in the virialized
region of the galaxy. In  Model A  the density peak  is nearly
uniformly distributed around the geometrical centre; it forms a
massive peak and several small sub-units. In Model B we note the
evolution of two distinct units, one evidently smaller than the
other, which eventually merge to form a single object; the merger is
completed at $ z \sim 1.3$.

Once the collapse has started, while the collisionless Dark Matter
begins to form clumps of high density the gas is decelerated  and
heated up by its own viscous friction which causes the onset of
shock processes; the kinetic energy of the gas is largely turned
into internal energy by such shocks; strong cooling begins to occur,
thus initiating star formation. Baryonic matter, in the form of
stars, can now freely fall in the potential wells of Dark Matter.

The first stars, in general in small isolated groups, start  to form
at $z \simeq 12$ in Model A and $z \simeq 15$ in Model B. Star
formation proceeds and as shown below it occurs in a sort of broad
burst whose duration is of the order of one or two Gyrs; then it
slows down until it (almost) stops.

Following the overall collapse, the energy injection by shocks  and
stellar feedback increases the energy (temperature) of the gas left
over by stars to the point that part of it can overwhelm the
gravitational potential. The gas escapes the galaxy and star
formation is eventually quenched off.

At the end of the galaxy and star formation period, the morphology
of all our models seems to nicely resemble that  of real galaxies.

To highlight the structure of our model galaxies, first we  analyse
their three-dimensional shape. To this aim we  chose a suitable
cut-off radius and consider the stellar and Dark components inside
this radius. The axial ratios can be roughly estimated as

\begin{equation}
\frac{b}{a} = \sqrt{\frac{\Sigma m_i y_i^2}{\Sigma m_i x_i^2}}\qquad \qquad
\frac{c}{a} = \sqrt{\frac{\Sigma m_i z_i^2}{\Sigma m_i x_i^2}}
\end{equation}

\noindent where $x, y, z$ are the coordinates along the axis of the
resulting ellipsoid. For Model A, inside the half mass radius,  we
find the axial ratios $b/a = 1.08$ (stars) and $1.14$ (Dark Matter),
and $c/a=1.07$ (stars) and $1.17$ (Dark Matter).  For Model B we get
$b/a=1.04$ (stars) and  $1.14$ (Dark Matter), and $c/a=1.00$ (stars)
and $0.96$ (Dark Matter). We can therefore state that, in general,
the stars are nearly spherically distributed, while the Dark Matter
haloes show a weak asymmetry. Owing to the smallness of the
deviation, the two models are considered  as spherically symmetric.

The panels of  Fig. \ref{vaucA_vaucC} show the final
\textit{surface} density profiles of the stellar contents of  Models
A and B projected onto the $xy$ plane when observed \textit{face
on}. The solid line are the \citet{Sersic68} surface density
profiles

\begin{equation}
\Sigma = \Sigma_e \exp \{-(0.324-2m)[(R/R_{eB})^{1/m} - 1 ] \}
\end{equation}

\noindent where $R_{eB}$, the \textit{effective radius}, is the
radius containing half mass of the galaxy (in this case referred to
cylindrical symmetry), $\Sigma_e$ is the mass density inside this
radius, and $m$ is a numerical parameter which correlates with the
galaxy absolute magnitude; for $m=1$ one finds the exponential law
of dwarf ellipticals, while for $m=4$ one gets the \citet{Vauc48}
law. The agreement between the numerical results and the de
Vaucouleurs law is remarkably good for Model A; only in the very
central regions (i.e. $r \le 2$ kpc) the density of star particles
seems to be too peaked. Model B seems well fitted by a Sersic law
with $m \sim 6$.   The best-fits have been made   minimizing the
dispersion of the  points $\sqrt{(\rho_i(r)-\rho_{fit}(r))^2}$) with
respect to the analytical relationship. As noticed by
\citet{Caon93}, one would actually expect smaller values of $m$ for
smaller galaxies, while  we obtain the opposite. This is a drawback
of the present models, which however should not be rejected because
of it. Profiles with m=6 and m=4 are indeed quite similar, so that
most likely the present simulations do not possess the resolving
power to distinguish between the two. It is safe to say that the
general trends of the mass profiles resemble those of real galaxies,
but a detailed comparison between theory and observations is not yet
possible with the present models.

Considering the models as roughly spherical, we  try to fit their
three-dimensional density profiles with theoretical analytical laws,
i.e. the \citet{Hernquist90} law for the stellar component

\begin{equation}
\rho(r) = \frac{M}{2 \pi} \frac{a}{r} \frac{1}{(r+a)^3}
\end{equation}

\noindent where $M$ is the total mass of the stellar system and $a$
is a scale length that depends on the half-mass radius $r_{1/2} =
(1+\sqrt{2})a$, and the \citet{NFW96} universal profile for the Dark
Matter component

\begin{equation}
\frac{\rho(r)}{\rho_{crit}} = \frac{\delta_c}{(r/r_s)(1+r/r_s)^2}
\label {nfw_1}
\end{equation}

 \noindent
where $\rho_{crit}$ is the critical density, $r_s$ is a scale radius
at which the slope of the density profile (\ref{nfw_1}) is $-2$, and
$\delta_c$ is a characteristic (dimensionless) density that depend
on the cosmological model and $r_s$ \citep[see][for details]{NFW96}.

Fig. \ref{densA_densC} shows the mass densities for our stellar and
Dark Matter components compared with the theoretical profiles {} (fits
are obtained the same way as above): the
agreement seems quite good in the middle regions. Obviously, little
can be said about the most central regions (i.e. less than $ \sim
2.5$ kpc), due to the effects of the softening length and, in
general, of numerical nature \citep[see e.g.][for a discussion of
these issues]{Power03}.

For both models the Dark Matter profile is very steep in the
innermost central regions. Current  cosmological simulations give
profiles with inner slopes between $ \sim -1$ \citep{NFW96} and $
\sim -1.4$ \citep{Moore98}. However, there are evidences that the
inclusion of non-adiabatic treatment of the gaseous component, and
particulary the star formation process, could steepen the Dark
Matter density profile \citep[e.g.][]{Lewis00,Kawata03,Gnedin04}.
\citet{Power03} also argue that poor temporal resolutions could give
rise to artificial cusps in the innermost regions. Therefore,
considering the uncertainty still affecting the whole issue, the
present results are acceptable.

Furthermore, we  note that, contrary to common expectations, in the
central regions of our models the density of Dark Matter exceeds
that of baryonic matter (stars): the Dark Matter halo seems to be
strongly concentrated. We adopt the definition of concentration by
\citet{NFW96} as the ratio $c=R_{vir}/r_{s}$ between the virial
radius $R_{vir}$ of the Dark Matter halo and the scale radius of
relation (\ref{nfw_1}) $r_{s}$. As virial radius we adopt the radial
distance at which the mean total density of the systems is 200 times
the background density; so, in this estimate both Dark and Baryonic
matter are included. However, owing to the small fraction of baryons
with respect to the total mass, the results does not significantly
depend on their presence. The scale radius is derived from the
best-fit of our numerical results to the NFW profile (see Fig.
\ref{densA_densC}). The concentrations we find for Models A and B
are $\sim50$ and $\sim 20$, respectively.

These values are higher than current expectations which indicate
concentration of $\sim 10$ for galactic haloes \citep{NFW96}. This
drawback of our models requires a careful future analysis. First of
all, as already mentioned, the presence of stars can steepen the
Dark Matter profiles. In this contest the effects of different star
formation recipes and/or different choices for the model parameters
should be investigated. Second, quasi-cosmological initial
conditions ignore the possible late-time infall of gaseous matter
from the IGM into the virialized halo of the new born galaxy.
Moreover, the observational estimate of the mass profiles in the
centre of galaxies is derived with the aid of (theoretical)
mass-to-light ratio for the stellar components, which depend on the
IMF as well as stellar evolution properties. While these latter are
somewhat established, this is not the case of the IMF. Passing from
one IMF to another may easily change the mass-to light ratios by a
large factor \citep{PortLarTan04}. In addition to this, our galaxies
seem to be completely relaxed and in equilibrium (see below). This
probably means that a baryon-dominated centre is not strictly
necessary to secure the stability of the systems. Last, the density
profiles of the luminous component of our models fairly agree with
those of \citet{Hernquist90}. Therefore, as far as we can tell,
neither theoretical nor observational evidences actually exist
constraining the central regions of galaxies to be \textit{always}
baryon-dominated, each galaxy being peculiar in its formation and
virialization history. In conclusion, we feel confident to consider
our models as reliable representations of the gross scale features
of real galaxies and the grand design of galaxy formation in which
these numerical simulations are framed.

Finally, the effective radius of the galaxy is calculated using the
Hernquist density profile, so that

\begin{equation}
R_{eB} \simeq \frac{R_{1/2}}{1.33},
\end{equation}

\noindent \citep{Hernquist90}. Table \ref{tabba} gives the values of
the parameters concerning the global properties of the density
profiles of the models.

%%%%%%%%%%%%%%%%%%%%%%%Table 3

\begin{table}
\centering \caption{Masses and half mass radii for the various
components of Models A and B. Masses are in units of
$10^{12}M_{\odot}$, radii are in kpc.}
\begin{tabular}{|l|l|l|}
\hline
Model & A & B \\
\hline
Final star mass $M_s$            & 0.091       &  0.0029 \\
\hline
Final gas mass (collapsed)       & 0.062       &  0.0004 \\
\hline
$M_s/M_{bar}$                    & 0.56        &  0.82   \\
\hline
Half mass radius of stars        & 7           &  1      \\
\hline
Effective radius of stars        & 5.2         &  0.8    \\
\hline
Sersic law index $m$             & 4           &  6      \\
\hline
Half mass radius of DM           & 52          &  15     \\
\hline
Concentration (NFW profile)      &  50         &  20     \\
\hline
Virial radius                    &  300        &  41     \\
\hline
Age of the last model (Gyr)      & 13          &  5      \\
\hline
\end{tabular}
\label{tabba}
\end{table}

\subsection{The Star Formation History} \label{ssss}

The history of the \textit{star formation rate}, in solar masses per
year, is shown in Fig. \ref{sfrA_sfrC} for Model A (top panel) and B
(bottom panel). They can be compared, for example, with those by
\citet{ChiosiCarraro02} and \citet{Kawata99,Kawata01b,Kawata01a}. It
is evident that in all cases the galaxy goes through a phase of
strong stellar activity, in a sort of burst-like mode lasting about
2 Gyr, followed by very slow activity (of the order of 1-5
$M_\odot$/yr) that continues for long periods of time. Remarkably
the bulk of stars are in place at the galaxy age of 2.5 Gyr, i.e. at
the redshift $z=1.95$ for Model A  and $z=2.2$ for Model B.

This behaviour shares the properties of the hierarchical and
monolithic scheme. It is hierarchical because in the proto-galaxy
lumps of matter merge together to assemble the  galaxy we see today.
The merging history is complete at redshift $z=2.5$ for the Model A
and $z=2.2$ for Model B, except one major merger which happens at $z
\sim 1.3$. It is monolithic because the largest fraction of the star
mass is built up very early before z=2, for all models.

The results we have obtained clearly favour the \textit{monolithic}
scheme of galaxy formation (cfr. Section \ref{intro}) as a unique
episode of star formation during very early stages can generate a
structure very similar to what we see today; even in the case of a
subsequent major merger as that of Model B, the star formation
history does not undergo substantial changes, perhaps due to the
small amount of gas left over by the first burst of activity.

How star formation actually works obviously depends  on the initial
conditions. In our models, the results favour the mergers of small
sub-units in very early epochs rather than the formation of big
galaxies from mergers of other galaxies at much later times.

 A huge fraction of the \textit{collapsed} gas is turned into
stars and also  a considerable amount of gaseous material is blown
away before it can collapse in the proto-galaxy potential well. With
the present recipes, the efficiency of gas restitution by dying
stars refuels the interstellar medium with sufficient new gas to
engender minimal star formation after the galaxy has reached fully
relaxed conditions. This is less of a problem here because the poor
mass-resolution makes the effects on the star formation rate of the
transformation of one or two particles per step more dramatic than
they actually are. Moreover, slightly different conditions on the
recipe for the star formation rate could get rid of it. In relation
to this, many experiments have recently been made to evaluate and
simulate the effects of stellar energetic feedback on the
interstellar medium \citep[cfr. for instance][]{Springel03,Marri03}.
Sufficiently strong feedback would stop the star formation activity
by evaporating the cold gas clouds in which new stars would form.
However  the poor resolution of the present numerical simulations
does not allow us to properly investigate this problem. As already
pointed out, we are not able to reproduce the effects of stellar
re-heating on the kinetics of gas particles, so that we must  limit
ourselves to just fuel the thermal budget, which actually proves to
be ineffective, due to the strong cooling phenomena which dominate
the gaseous non-adiabatic processes at the  temperatures we are
dealing with. Therefore, past the strong, dominant episode of star
formation, the long tail of stellar activity can be simply ignored
for all practical purposes, for instance the derivation of
spectro-photometric properties.

As already mentioned, the by far dominant episode of star formation
occurred at very early epochs and  lasted for a significant fraction
of the Hubble time. The duration of it (about 2 Gyr) is longer in
Model B than in Model A thus leading to a higher ratio between the
total baryonic mass and the left-over gas mass (cfr. Table
\ref{tabba}). All this bears very much on the chemical evolution of
the (model) galaxies as discussed in Section \ref{jazz}.

Table \ref{tabba} lists the mass of the various components at the
end of the simulations together with the half mass radius of the
stellar and Dark Matter distributions. CDM turns out to be strongly
concentrated, as already noticed. The final stellar-to-gas ratios
(considering only the collapsed and/or reprocessed gas) turns out to
be $\sim 1.47$ and $\sim7.25$ for Models A and B, respectively; as
already pointed out, Model B results much more efficient in
forming stars, leaving only $\sim 13 \%$ of the baryonic
\textit{collapsed} particles in the gaseous form (see Table
\ref{tabba} for details). The mass of Model B is not small enough to
consider it a \textit{dwarf} elliptical, but this results is still a
bit surprising, since we would expect that a less deep potential
well would produce less efficient star formation processes
\citep[see e.g.][]{ChiosiCarraro02}. Remarkably, the vast majority
of stars in Model B are formed \textit{before} the major merger at
$z\sim1.3$, thus ruling out the possibility that the merger could be
responsible for the different efficiency in forming stars (indeed, a
small increase in star formation can be noticed in coincidence with
the merger, but \textit{after} the formation of the bulk stellar
content stars).  A possible explanation can be sought   in the
presence of \textit{two} distinct potential wells. The central
regions of an even small potential well can indeed be more efficient
in forming stars, than the peripheral regions of a bigger one. In
other words, while the gaseous particles in the outskirts of a
single, deep potential well begin to  heat up by stellar winds and
supernova explosions before they can fall inside the well and
therefore are likely subtracted from the star forming activity, in
presence of two (or more) potential wells a larger percentage   of
gas particles, in  different regions of the forming galaxy, can be
turned into stellar particles.

Fig.\ref{ageradA_ageradC} shows the age of the stars as a function
of their radial position for Models A and B.  It clearly illustrates
how a galaxy is assembled. The star formation process in the central
regions goes on for very long periods of time; younger and more
metal-rich  stars are therefore expected to be found in the core of
galaxies, whereas older and more metal-poor stars are likely to be
found in the outer regions.

\subsection{Energy}
The panels of Fig.\ref{eneA_eneB}, from top to bottom, show  the
variation as a function of time  of  the kinetic, thermal, total and
potential energies (all expressed in the physical units we have
adopted in the code) of Model A.  The same quantities for   Model B
have similar behaviour. The trends resemble nicely what we expect to
see in an expansion and subsequent collapse of an isolated system;
notice that the thermal energy, which is provided only by the gas,
is very small when compared to the kinetic and the potential terms.
The total energy is not strictly conserved (only a few per cent), as
the inclusion of non-adiabatic processes such as cooling causes the
loss of part of the energetic budget of the gaseous component.

Fig. \ref{virA_virC} shows the  Virial Trace $V=-2E_{Kin}/E_{pot}$
as a function of time for the two models. It is easy to follow the
relaxation of the system, which tends to a state of equilibrium
after some small oscillations (if the systems had been strictly
collisionless and more symmetrical these oscillation would have been
much stronger, as we expect in a \textit{violent relaxation}
process). The expected value for an ideal system would be $V=1$, but
here the presence of viscosity and cooling of the gas tend to shift
the ratio to somewhat lower values to the loss of energy. It is
worth noticing in Model B the bump in coincidence with the major
merger occurring at $\sim 3.5$ Gyr.

Moving now to heating processes, the rate of supernova explosions of
Model A  is shown in Figs. \ref{snIA_snIIA}  (the trend is nearly
the same for Model B). The top and bottom panels are for Type IA and
Type II supernovae, respectively. As expected the rate of Type II
supernovae strictly follows the star formation rate: as massive
stars are short lived, as soon as star formation starts and/or
ceases the same does the Type II supernova rate. The case of Type Ia
is different because they require a certain amount of time to show
up and continue to explode for very long time after the end of the
star forming period.

\subsection{Metallicity} \label{jazz}

 Fig. \ref{metA_metC} displays the evolution of the mean
metallicity Z and mass abundance of Fe  and O both for  stars and
gas in Models A and B; the ratio [O/Fe] is in the standard
spectroscopic notation. The plots can be compared with those in
Fig.5 of \citet{ChiosiCarraro02}.

Looking at the [O/Fe] vs age relation displayed in Fig.
\ref{metA_metC}, we notice that the so-called $\alpha$-enhancement
problem is present. Both models have the stellar content strongly
enhanced in $\alpha$-elements, however the degree of enhancement is
lower in Model B (lower total mass) and in both cases decreases with
the age. These trends are agree with the observational information,
see the review by \citet{Chiosi00}, and  the  similar analysis made
by \citet{Tantalo02} using the \citet{ChiosiCarraro02} models.

The final mean stellar metallicity in Model A is nearly solar ($<Z>
\simeq 0.8 Z_{\odot}$). Remarkably also Model B shows similar values
of mean metallicity even if its evolution was stopped much before
the present age, thus showing how the chemical enrichment seems to
be almost completed in the first Gyr of  a galaxy life.

Fig. \ref{chimbinA} shows the final number distribution of stars per
metallicity bin (Z/Z$_\odot$), limited to Model A: $4\%$ of stars
have mean total metallicity lower than 1/3 solar, $67\%$  have mean
total metallicity in the range 1/3 solar to solar, a significant
fraction goes up to three times solar, and a tiny fraction even up
to 5 times solar. The low fraction with metallicity lower than 1/3
solar secures that the so-called G-dwarf problem does not occur,
whereas the tiny fraction around 5 times solar could explain the UV
excess in elliptical galaxies. See  \citet{Bressan94} for a thorough
discussion of these topics.

At any age a large scatter in the metallicities of the star
particles has to be expected. This is shown in Fig.
\ref{timechimA_timechimC}. The situation resembles the one observed
among the stars of the Galactic Disk
\citep{Edvardsson93a,Edvardsson93b}.

Finally, in Fig. \ref{radmetA} we display the spherically averaged
radial gradient in metallicity for Model A and B; stars closer to
the centre are also more metal rich. For elliptical galaxies,
\citet{Davies93}  find the metallicity gradient $\Delta log(Z) /
\Delta log(r) \simeq -0.2 \pm 0.1$. The gradient of Model A  is
$\simeq -0.30$, in good agreement with the observational  estimate.
Model B shows instead a steeper gradient, $\simeq -0.50$, due to the
larger amount of stars produced towards the centre.

As a last remark we like to report here that similar simulations
with almost identical initial conditions but with the
\citet{Arimoto87} IMF yield very high values of the final mean
metallicities, i.e. up to 5 or 6 times the solar abundance
\citep{Merlin05}. The reason is that the \citet{Arimoto87} IMF
favour massive stars and Type II supernovae and therefore predict
much higher rates of gas restitution than the \citet{Kroupa98} or
\citet{Salpeter55} IMF \citep[see e.g.][]{Lia02}. As such high mean
metallicities do not agree with current observational estimates, in
the present study we prefer to abandon the \citet{Arimoto87} IMF
even if it was originally tailored to model elliptical galaxies. The
recent calculation of stellar yields by \citet{Portinari98} which
are incorporated in the \citet{Lia02} algorithm for chemical
enrichment in N-Body PD-TSPH simulations, show that IMFs too skewed
toward massive stars are not required. The reader should refer to
\citet{Merlin05} for all details about these models with
\citet{Arimoto87} IMF.

\subsection{Galactic Winds}

As already reported in Section \ref{intro} elliptical  galaxies
ought to suffer galactic winds in order to explain the
colour-magnitude relation. To investigate this important issue,  in
Fig. \ref{tempdensA_tempdensC} we show the temperature vs density
relationships for all gas particles of our models. The gas left over
by star formation is made of four components - hot and cold,
chemically enriched and unprocessed \citep[cfr.][]{ChiosiCarraro02}.
The hot gas, which is the dominant component, has relatively high
densities: heated up by energy injection is about leaving the
potential well of the galaxy. The fact the hot gas constitutes
almost the totality of the gas content is due to the efficiency of
the energy feed-back and the generally large total mass of the
systems (relatively deep potential wells) which inhibits the
occurrence of earlier galactic winds
\citet[cfr.][]{ChiosiCarraro02,Kawata03}. Note that in Model B, as
fewer gas particles are left over by star formation, the situation
is less defined then in Model A, where we can clearly see the final
dependence, for the hot component, of temperature on density.

Fig. \ref{winds} shows the radial velocity of the gas particles
together with the radial profile of the escape velocity. The amount
of gas (excluded the fraction which did not take part to the
collapse of the proto-galaxy) which, at the end of our  simulations,
has the radial component of its velocity greater then the escape
velocity at its distance from the centre of mass of the system, can
freely leave the potential well of the galaxy and can be considered
as gaseous mass ejected as galactic wind. In all simulations there
is a layer beyond which all the gas particles meet these
requirement; the percentage of the gas which can be considered as
galactic wind is $\sim 21\%$ (Model A) and $\sim 58\%$ (Model B) on
the total of collapsed gas. Note that these percentages are very
different from one case to another; this could be due to several
factors, among which  the different star formation history (as
already pointed out, star formation is much more efficient in Model
B than Model A, thus leaving a smaller amount of gas in the central
regions of the system), and the different masses of the systems,
since galactic winds are expected to be more efficient in galaxies
of low mass \citep{ChiosiCarraro02}. At large distances from the
galaxies, some particles with almost zero radial velocity can be
found (see Fig. \ref{winds}). They are the un-collapsed gas
particles that did not take part to the galaxy formation process.
They  cannot be considered as  galactic wind as they have not been
blown away from the galaxies.

Finally, we would like to note that a better  treatment of the
gaseous component should be considered such as the multi-phase
scheme of \citet[e.g.][]{Marri03} that would allow us to describe in
a more realistic fashion  both chemical enrichment and galactic
winds.

\section{Conclusions}\label{conclusions}

In this study, with the aid of N-body simulations based  on
quasi-cosmological initial conditions, we have followed the
formation and evolution of two early-type galaxies from the stage
when they separate from global expansion of the Universe to their
collapse to virialized structures, the formation of stars and
subsequent nearly passive evolution up to the age of 13 Gyr. The
cosmological background is the Standard CDM. We have highlighted the
structural, dynamical and chemical properties of the models to be
compared with observational data.

The models contain several parameters, among which we recall

\begin{itemize}

\item The softening parameter for Dark Matter, gas and stars;

\item The specific efficiency  $c_*$ driving the star formation rate;

\item The constant $k_{SN}$
defining the fraction of energy released  by a supernova explosion
which goes into the gas as thermal energy.
\end{itemize}

The uncertainty on these  parameters opens the gate to future
studies; better \textit{fine tuning} of these quantities is required
to get results in very close agreement with the observational data.
Nevertheless the results we have obtained are satisfactory and in
agreement with other studies of the same subject.

Two systems made of Dark and baryonic matter, in current
cosmological proportions, with positions and velocities for both
type of particles derived from the spectrum of cosmological
perturbations suited to the cosmological model under consideration
have been investigated.

All models started from the conform expansion of the Hubble flow,
with the lowest rigid body rotation allowed by the CDM Universe
\citep{Barnes87}. They all have been followed for a large fraction
of the Hubble time, through the phases of \textit{turn around} and
subsequent collapse of the galactic proto-clouds of gas to small
\textit{clumps} caused by the initial density perturbations to  the
formation of equilibrium structure made of Dark Matter, stars and
remaining gas. Hydrodynamical and thermal processes caused by
\textit{cooling}, star formation, chemical enrichment, and
\textit{heating} have been included and followed in detail.

The final systems morphologically resemble real elliptical galaxies
of intermediate dimensions; this is proved by their surface density
profiles matching the theoretical ones by \citet{Vauc48} and/or
\citet{Sersic68} that are expected for galaxies of the same mass and
scale lengths. \citet{Hjorth91} have indeed showed that
\textit{violent relaxation} in a deep potential well leads to
systems obeying the  de Vaucouleurs  law. \citet{KatzGunn91} and
\citet{Kawata99} outline that dissipative processes,  followed by
the \textsc{PD-TSPH} code in great detail, bear very much on the
final results.

Star formation lasts longer in the central regions than in the
outskirts thus producing gradients in metallicity that closely agree
with the observational data for elliptical galaxies
\citep{Davies93}.

It is woth stressing that in any case
the models still need to be improved in several aspects: (i)
First the star formation that even if shaped in a huge initial burst
followed by quasi-quiescence, never completely stops, because of
poor mass resolution and/or still inadequate prescription for the
physical conditions at which stars can be formed. The long tail of
minimal stellar activity does indeed affect the spectro-photometric
properties (e.g. colours) of the model galaxies that would disagree
with observational data. (ii) Second, the density profiles that are
not fully satisfactory, as in the central regions the baryons  do
not dominate the spatial density of matter as suggested by the
observational data. (iii) Third, there's no right correspondence
between the values of Sersic parameter $m$ to galaxies of different
mass and the those derived from observational data. (iv)  Finally,
the predicted mass-metallicity relation as compared to the one
inferred from observational data (colours, e.g. the CMR). The
present model B with lower mass actually suffers more star formation
and metal enrichment in turn. The reason for that is not yet clear.
The present results cannot yet recover the metallicity-mass
relationship obtained by \citet{ChiosiCarraro02} that fairly agreed
with the observational information.

Nevertheless, it is worth noticing  that our main results are in
good agreement with those by \citet{Kawata99}, who used the same
method to generate the initial conditions, as well as with those  by
\citet{Kawata03} and \citet{Kobayashi05}, who instead used the more
refined method otherwise known as  \textit{resimulation technique}
(see Section. \ref{comp_ini_cond}). This  point is particularly
relevant as it shows that our approach, developed independently and
in parallel to that of \citet{Kobayashi05}, can be considered
sufficiently accurate.

Our simulations  support the \textit{revised monolithic scenario} of
galaxy formation. Through the hierarchical collapse of Dark Matter
haloes, clouds of gas meet, heat up, cool and eventually collapse
forming stars in a single episode of star formation occurred in the
far past, during the virilization of the proto-galaxy, perhaps
followed by minimal stellar activity in the central regions and
almost negligible morphological and structural evolution.

%=============== ACKNOWLEDGEMENTS ======================%
\begin{acknowledgements}
The authors would like to thank Laura Portinari, Bepi Tormen,
Mauro D'Onofrio and Luigi Secco for the many helpful discussions and
constructive criticism,  Lorenzo Piovan for carefully reading the
manuscript, and the anonymous referee for the many useful
suggestions. This study has been financed by the University of
Padua under the special contract "Formation and evolution of
elliptical galaxies: the age problem".
\end{acknowledgements}

%=================== BIBLIOGRAPHY ======================%

%\bibliographystyle{apj}           % Files .bst

%\bibliography{mnemonic,biblio}    % Files .bib

%=================== END DOCUMENT ===================%

\end{document}